\documentclass[onecolumn,sort&compress,numbers]{els-mrw} 

\usepackage{amsmath,amssymb,amsfonts,amsthm,makeidx,graphicx}
\usepackage{txfonts}
\usepackage{helvet}
\usepackage{dsfont}
\usepackage{slashed}
\usepackage{setspace}

\usepackage[colorlinks=true,linkcolor=blue,citecolor=blue,urlcolor=blue]{hyperref}

\newcommand{\conjg}[1]{\ensuremath{\hspace{1pt}\overline{\hspace{-1pt}#1\hspace{-1pt}}}\hspace{1pt}}

        \def\longlonglongrightarrow{
        \relbar\joinrel\relbar\joinrel\relbar\joinrel\relbar\joinrel\relbar\joinrel\relbar\joinrel\rightarrow}

\def\mD{\ensuremath{\mathcal{D}}}

\def\mS{\ensuremath{\mathcal{S}}}

\begin{document}


\chapter{Hadron physics with functional methods}\label{chap1}

\author[1]{Gernot Eichmann}%
%

\address[1]{\orgname{University of Graz, NAWI Graz}, \orgdiv{Institute of Physics}, \orgaddress{Universitätsplatz 5, 8010 Graz, Austria}}

\articletag{Chapter Article tagline: update of previous edition, reprint.}

\maketitle

\begin{abstract}[Abstract]
	We give a pedagogical introduction to hadron spectroscopy and structure studies using functional methods.
    We explain the basic features of Dyson-Schwinger, Bethe-Salpeter and Faddeev equations, which are employed
    to calculate the spectra of mesons, baryons and four-quark states.  
    We discuss dynamical mass generation as a consequence of the spontaneous breaking of chiral symmetry,
    which is intertwined with the emergence of the light pions as Goldstone bosons of QCD.  
    We highlight the importance of diquark correlations in the baryon sector,
    while for four-quark states such as the light scalar mesons and heavy exotics the dominant two-body clusters are typically mesons.
    We conclude with a brief discussion of hadron matrix elements like electromagnetic form factors and how 
    vector-meson dominance is an automatic outcome of functional equations. 
\end{abstract}

\begin{keywords}
 	Dynamical mass generation \sep Dyson-Schwinger and Bethe-Salpeter equations  \sep Diquarks \sep Exotic hadrons \sep Form factors
\end{keywords}


\begin{glossary}[Nomenclature]
	\begin{tabular}{@{}lp{34pc}@{}}
        axWTI & Axialvector Ward-Takahashi identity\\
        BSE & Bethe-Salpeter equation \\
        DSE & Dyson-Schwinger equation \\
        GMOR & Gell-Mann-Oakes-Renner \\
		LHC & Large Hadron Collider\\
        PCAC & Partially conserved axialvector current \\
        PDG & Particle Data Group \\
        QCD & Quantum chromodynamics \\
		QFT & Quantum field theory
	\end{tabular}
\end{glossary}

\newpage

\section*{Objectives}
\begin{itemize}
	\item Understand how dynamical mass generation comes about and why the pion is so light.
	\item Explain the basic concepts behind Dyson-Schwinger, Bethe-Salpeter and Faddeev equations.
	\item Overview of baryon spectroscopy with three-quark and quark-diquark calculations.
	\item Exotic four-quark spectroscopy and the dynamical generation of resonances.  
    \item Brief overview of hadron structure calculations with functional methods.
\end{itemize}

\section{Introduction}\label{intro}

The spectrum and structure of hadrons encodes a wealth of interesting phenomena.
Hadrons are composites of quarks and gluons, the fundamental degrees of freedom in Quantum Chromodynamics (QCD) which is
the theory of the strong interaction. Because quarks and gluons carry color,
they are confined in hadrons and cannot be directly probed in experiments.
Traditionally, hadrons are classified in mesons as quark-antiquark ($q\bar{q}$) states
and baryons as composites of three valence quarks ($qqq$). 
However, nowadays we know that the situation is substantially more complex: 
There is experimental evidence for exotic hadrons like tetraquarks ($qq\bar{q}\bar{q}$) and
pentaquarks ($qqqq\bar{q}$), and there could also be
hybrid mesons with extra valence gluons 
or  glueballs made of gluons only. 
Even our understanding of the nucleon  is far from complete: Protons ($uud$) and neutrons ($udd$)
consist of three light up and down quarks, whose masses  sum up to only about $1\%$ of the nucleon mass.
Thus, the overwhelming remainder must be  generated in QCD. This property of dynamical mass generation has to do with chiral symmetry,
which is also responsible for the small masses of the pions as the lightest mesons.

Even though QCD is fully defined by its Lagrangian, 
extracting observables is much more complicated
compared to the electroweak sector of the Standard Model.
For example, suppose we want to calculate the scattering of two electrons (Møller scattering) in Quantum Electrodynamics.
To do so, we just need to sum up the Feynman diagrams contributing to the process in the lowest orders of 
the electromagnetic coupling $\alpha_{em} = e^2/(4\pi) \approx 1/137$
and compare the resulting cross section with experiment.
In QCD, this strategy does not work for two reasons. 
First, confinement means that a process like $q\bar{q} \to q\bar{q}$ scattering is not directly measurable, so
one must find a passage to channel such processes to the observable world of hadrons.
Second, perturbative expansions are only possible at large momenta, where the strong coupling $\alpha_s$ is  small.
At small momenta, $\alpha_s$ becomes large and perturbation theory is no longer applicable to calculate hadron properties ---
instead, one needs nonperturbative methods.

\begin{wrapfigure}[22]{r}{0.45\textwidth}
\vspace{-0mm}
\includegraphics[width=\linewidth]{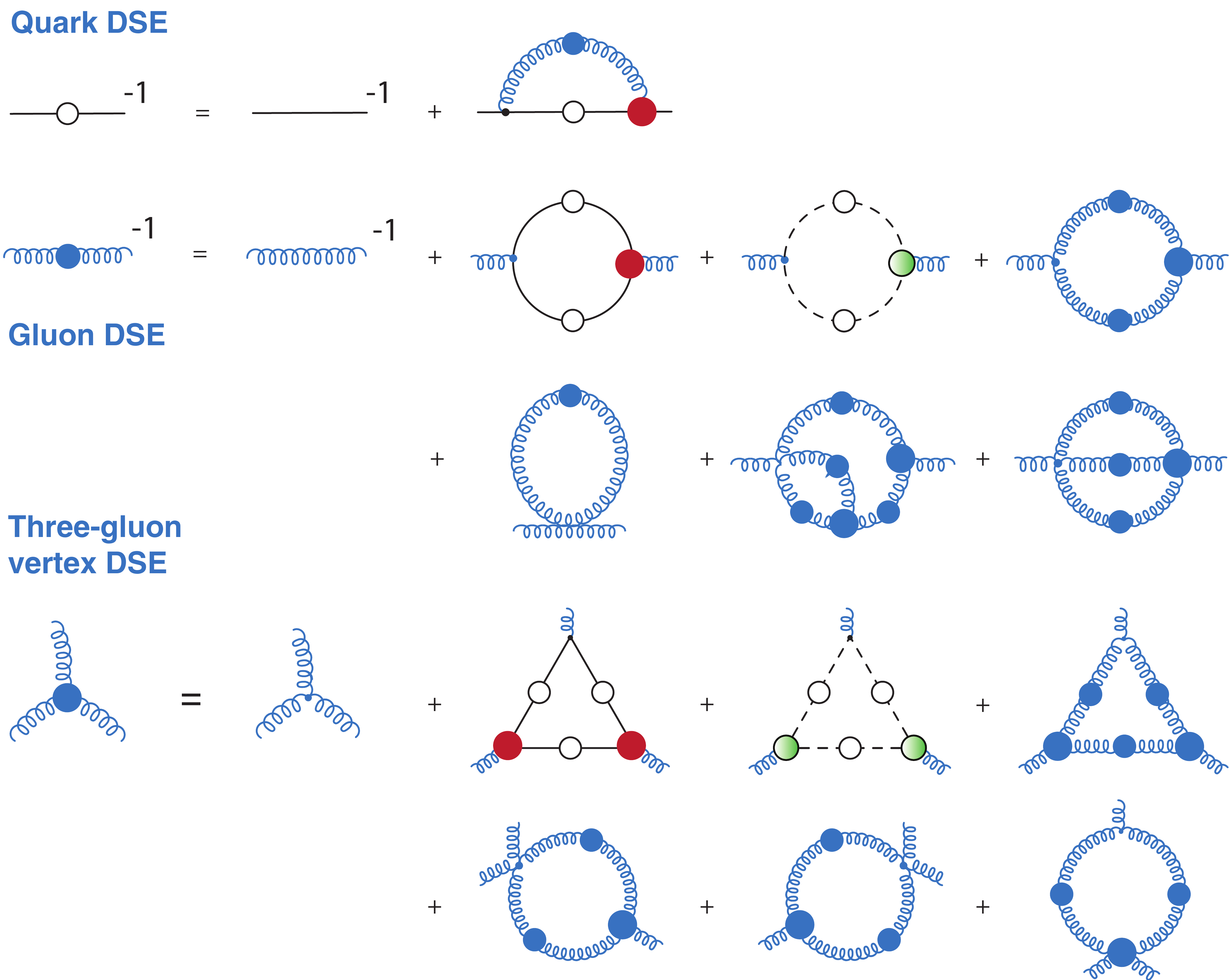} 
\caption{Coupled Dyson-Schwinger equations for the quark propagator, gluon propagator and three-gluon vertex.}
\label{fig:dses}
\end{wrapfigure}

Nowadays, lattice QCD has become the primary tool to calculate hadron properties from first principles.
There, one computes  correlation functions 
from the QCD path integral.
These have poles in momentum space corresponding to hadrons with masses $m_i$.
The Fourier transform of  a  pole yields an exponential falloff $\sim e^{-m_i\tau}$ in Euclidean time $\tau$, 
from where one can extract the hadron mass $m_i$ (see e.g.~\cite{Gattringer:2010zz}).

An alternative approach, which is also based on the path integral, is the one via functional methods. By taking functional derivatives of the path integral,
the classical equations of motion (Klein-Gordon, Dirac, Maxwell equations) are converted into quantum equations of motion,
which are the Dyson-Schwinger equations (DSEs)~\cite{Roberts:1994dr,Alkofer:2000wg,Bashir:2012fs,Eichmann:2016yit,Huber:2018ned,Fischer:2018sdj}. These are  coupled integral equations 
which relate QCD's $n$-point correlation functions with each other.
Fig.~\ref{fig:dses} shows exemplary DSEs for the two-point functions like the quark propagator and gluon propagator,
three-point functions like the three-gluon vertex, and the list goes on.
A similar method is the functional renormalization group, which leads to coupled integro-differential equations for the  $n$-point functions~\cite{Berges:2000ew,Pawlowski:2005xe,Dupuis:2020fhh}.
So, instead of computing correlation functions directly from the path integral like in lattice QCD, one derives relations from the path integral
to calculate them \textit{from each other}. While the Lagrangian defines the classical field theory,
the set of all $n$-point correlation functions defines the corresponding quantum field theory (QFT) completely.

Hadrons appear as poles in QCD's $n$-point functions. For example, the process $q\bar{q} \to q\bar{q}$ contains all possible meson poles.
This is how hadron masses are extracted in lattice QCD, and the same is true for functional methods.
As we will see in the following sections, one can  derive Bethe-Salpeter equations (BSEs) which are valid at these poles.  
These are the QFT analogues of the Schrödinger equation in quantum mechanics. 
The $n$-point correlation functions then appear in the kernels of these equations.
From a BSE one can calculate the mass of a hadron and its `wave function' (the Bethe-Salpeter amplitude). 
Once the wave function is known, one can compute hadron matrix elements like form factors, scattering amplitudes and parton distributions.

In the following we  discuss the basics of hadron spectroscopy and structure calculations with functional methods,
along with a brief survey of some results obtained so far.
We will frequently refer to the reviews~\cite{Eichmann:2016yit,Eichmann:2020oqt}, where details and references to the literature can be found.
We use Euclidean conventions throughout, which are collected in Appendix A of Ref.~\cite{Eichmann:2016yit}.

\newpage

            \begin{figure*}[t]
                    \centering
                    \includegraphics[width=1\textwidth]{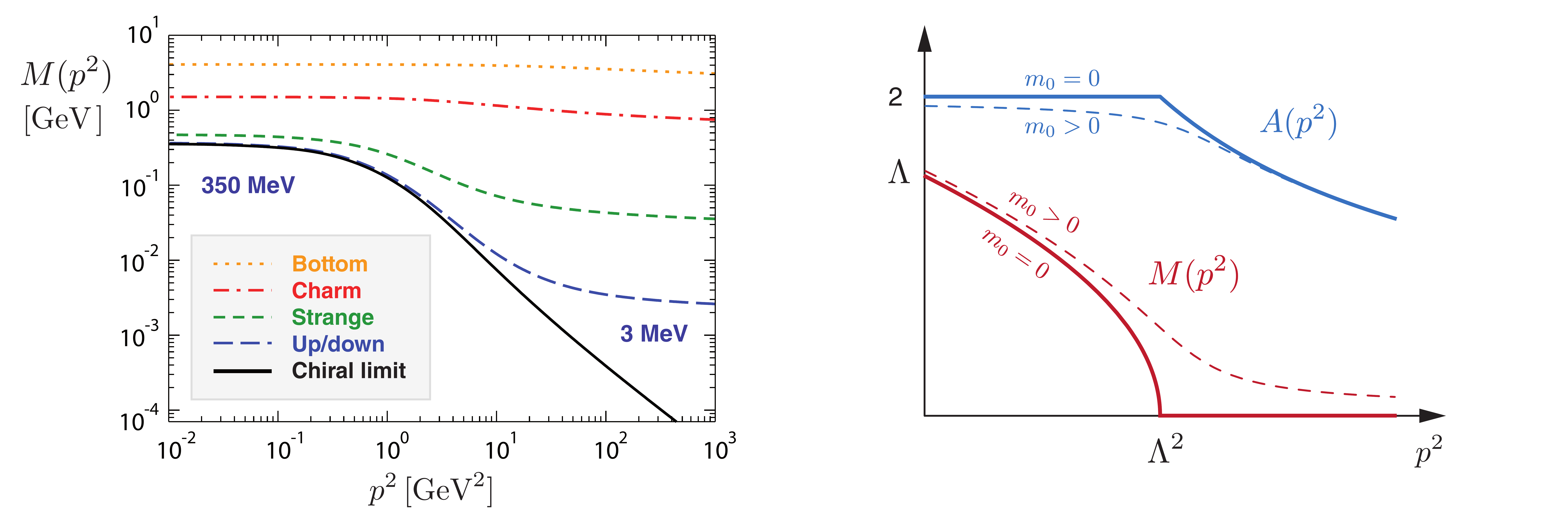}
                    \caption{Left: Quark mass function for different  flavors obtained by solving the quark Dyson-Schwinger equation~\cite{Eichmann:2016yit}. Right:
                             Dressing functions of the quark propagator in the Munczek-Nemirovsky model, Eqs.~(\ref{mn-sol1}--\ref{mn-sol2}). }
                    \label{fig:quark-mf}
            \end{figure*}

\section{Quarks}\label{sec:quarks}

  \subsection{Dynamical mass generation}\label{sec:dyn-mass}

  Light quarks are quite special because most of their mass is dynamically generated in QCD through their interactions with gluons. 
  The light up and down quarks have current-quark masses of only about 2--5 MeV, which arise from the Higgs mechanism as external inputs to QCD.
  In contrast, the mass of a proton composed of $uud$ is 938 MeV. This striking difference implies that nearly all the mass of the proton -- 
  and consequently the masses of nuclei, atoms, and the baryonic matter in the universe -- must be generated within QCD itself.
  The phenomenon is called dynamical mass generation, and it can be understood through the dynamical breaking of chiral symmetry, a nonperturbative effect arising from 
  the interactions of quarks and gluons which cause the quarks to gain mass.

  Dynamical mass generation is  visible in the quark propagator (among many other quantities), which is the two-point function 
  \begin{equation}\label{quark-general-0}
      S_{\alpha\beta}(p) = \int d^4x\,e^{-ipx}\, \langle 0 \,| \,\mathsf{T}\,\psi_\alpha(x)\,\conjg \psi_\beta(0)\,|\, 0 \rangle 
  \end{equation}
  and therefore the basic object describing a quark in QCD. Here, $\psi_\alpha$ and $\conjg \psi_\beta$ are the quark and antiquark fields,
  and $\mathsf{T}$ stands for time ordering.
  Because quarks are spin-$1/2$ particles, the most general  form of the quark propagator $S(p)$ in momentum space according
  to Lorentz invariance is
  \begin{equation}\label{quark-general}
      S(p)^{-1} = A(p^2)\left(i\slashed{p} + M(p^2)\right) \quad \Leftrightarrow \quad
      S(p) = \frac{1}{A(p^2)}\,\frac{-i\slashed{p} + M(p^2)}{p^2 + M(p^2)^2} = -i\slashed{p}\,\sigma_v(p^2) + \sigma_s(p^2)\,.
  \end{equation}
  It depends on two dressing functions,
  $A(p^2)$ and the quark mass function $M(p^2)$,
  or equivalently the two dressing functions $\sigma_v(p^2)$ and $\sigma_s(p^2)$ as combinations of them.
  The variable $p^2$ can take any value $p^2 \in \mathds{C}$,
  but in the following we are mainly interested in spacelike momenta $p^2 \in \mathds{R}_+$.
  For a free spin-$1/2$ particle with mass $m$, the above would simplify to $A(p^2) = 1$, $M(p^2) = m$ and therefore
  \begin{equation}\label{free-propagator}
     S(p) = \frac{-i\slashed{p} + m}{p^2+m^2} \quad \Leftrightarrow \quad
     \sigma_v(p^2) = \frac{1}{p^2 + m^2}\,, \quad
     \sigma_s(p^2) = \frac{m}{p^2+m^2}\,.
  \end{equation}

  Switching on interactions, the masses and couplings in a quantum field theory become momentum-dependent.
  As shown in the left panel of Fig.~\ref{fig:quark-mf}, the effect is quite drastic
  for the quark mass function $M(p^2)$:
  At large momenta, $M(p^2)$ coincides with the current-quark mass $m$, while at small momenta
  it is much larger by several hundred MeV and thereby defines a `constituent-quark mass', which is the relevant
  mass scale for the proton and other hadrons.
  
  That this effect is due to the spontaneous breaking of chiral symmetry can be seen as follows.
  In the chiral limit, where the current-quark masses vanish, the QCD Lagrangian has a chiral symmetry 
  $SU(N_f)_L \times SU(N_f)_R \simeq SU(N_f)_V \times SU(N_f)_A$, where $N_f$ denotes the number of quark flavors. 
  This is larger than the usual (vector) flavor symmetry $SU(N_f)_V$, which would be satisfied if all  quark masses were identical.
  Collecting all quark flavors in an $N_f$-dimensional vector $\psi$, the symmetry operations read
  \begin{equation}\label{va-tf}
      SU(N_f)_V \!\!: \quad \psi' = e^{i \sum_a\varepsilon_a \mathsf{t}_a}\,\psi\,, \qquad \qquad
      SU(N_f)_A\!\!: \quad \psi' = e^{i \gamma_5 \sum_a\varepsilon_a \mathsf{t}_a}\,\psi\,,
  \end{equation}
  where the $\mathsf{t}_a$ are the $SU(N_f)$ generators ($a=1 \dots N_f^2-1$) and the $\varepsilon_a$ are the parameters of the global transformations.
  The axial  $SU(N_f)_A$ symmetry only holds in the chiral limit and is explicitly broken if the quark masses are non-zero.
  Now, if chiral symmetry were preserved also at the quantum level, then all $n$-point correlation functions would be chirally symmetric.
  For the quark propagator this implies $\{ \gamma_5, S(p) \} = 0$, which follows from plugging in the axial transformation~\eqref{va-tf} 
  of the quark fields into Eq.~\eqref{quark-general-0}.
  Using the decomposition~\eqref{quark-general}, this yields $M(p^2)=0$, in contrast to the result shown in Fig.~\ref{fig:quark-mf}.
  Apparently there must be some dynamical mechanism that generates a quark mass `out of nothing', i.e., even in the chiral limit of vanishing current-quark masses.
  
\begin{figure*}[t]
                    \begin{center}
                    \includegraphics[width=0.9\textwidth]{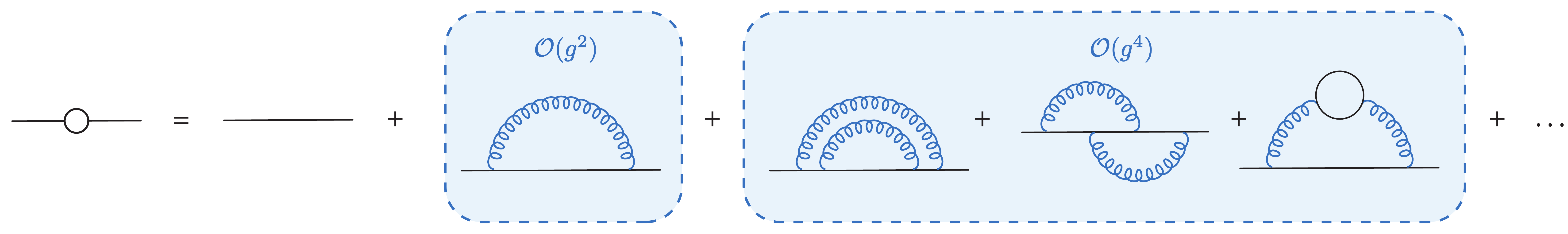}
                    \caption{Perturbative expansion of the quark propagator.}\label{fig:quark-perturbative}
                    \end{center}
                    \vspace{-3mm}
            \end{figure*}

  The fact that this effect is \textit{nonperturbative} means that one cannot produce it at any order in perturbation theory,
  which is the standard tool to make a QFT useful in practice. 
 When expanding the quark propagator in powers of $g$, as shown in Fig.~\ref{fig:quark-perturbative}, one generates all possible
  Feynman diagrams, where each comes with a power in the strong coupling $\alpha_s = g^2/(4\pi)$.
  If the coupling is small enough, one can stop the series after a few terms.
  For QCD this only works at large momenta, where $\alpha_s$ is small;
  this is what allows one to describe high-energy scattering processes.
  However, for small momenta $\alpha_s$ becomes large and perturbation theory fails. 
  The ingredients of the perturbative expansion are the tree-level quark-gluon vertex $\sim i\gamma^\mu$ and the tree-level quark propagator,
  which in the chiral limit ($m=0$) via Eq.~\eqref{free-propagator} is given by $-i\slashed{p}/p^2$ (up to a renormalization constant).
  Now, because both quantities contain one $\gamma-$matrix, 
   every possible perturbative diagram contains an odd number of $\gamma$-matrices.
  Thus, the trace of the quark propagator vanishes to all orders in perturbation theory!
  However, by Eq.~\eqref{quark-general} that trace is proportional to the quark mass function, which therefore vanishes in perturbation theory.
  
  How can we generate a mass nonperturbatively?
    A simple analogue of the perturbative expansion is the geometric series 
  \begin{equation}\label{geom-series-0}
     \sum_{n=0}^\infty x^n = \frac{1}{1-x} \qquad \text{for} \quad |x| < 1 \,,
  \end{equation}
  which  converges only for $|x|<1$.
  But suppose someone \textit{gave} you the `nonperturbative' equation
  \begin{equation}\label{geom-series}
     f(x) = 1 + x f(x) \quad \Leftrightarrow \quad f(x)^{-1} = 1-x\,,
  \end{equation}
  which has the solution $f(x) = 1/(1-x)$ for any $x$ except $x=1$.
  When inserting the l.h.s. into the r.h.s., then
  \begin{equation}\label{geom-series-3}
     f(x) = 1 + x f(x) = 1 + x + x^2 f(x) = 1 + x + x^2 + x^3 f(x) = \dots
  \end{equation}
  is still `nonperturbative' and exact at every step. In the geometric series, on the other hand, we dropped the last term which otherwise pulls the result back
  even if $x$ becomes large, and this is what makes the perturbative expansion fail.

\begin{figure*}[t]
                    \centering
                    \includegraphics[width=0.55\textwidth]{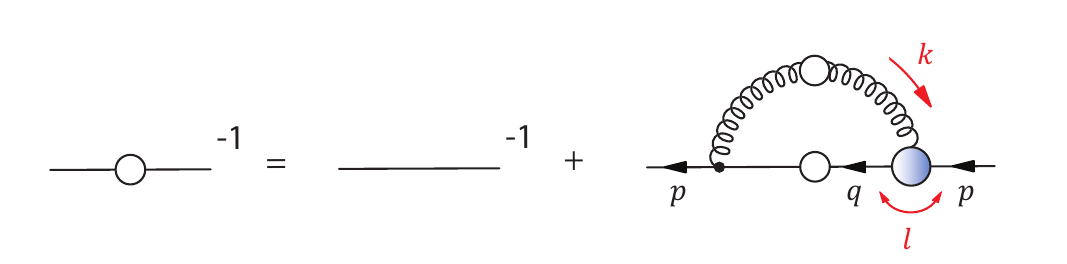} \hspace{5mm}
                    \includegraphics[width=0.4\textwidth]{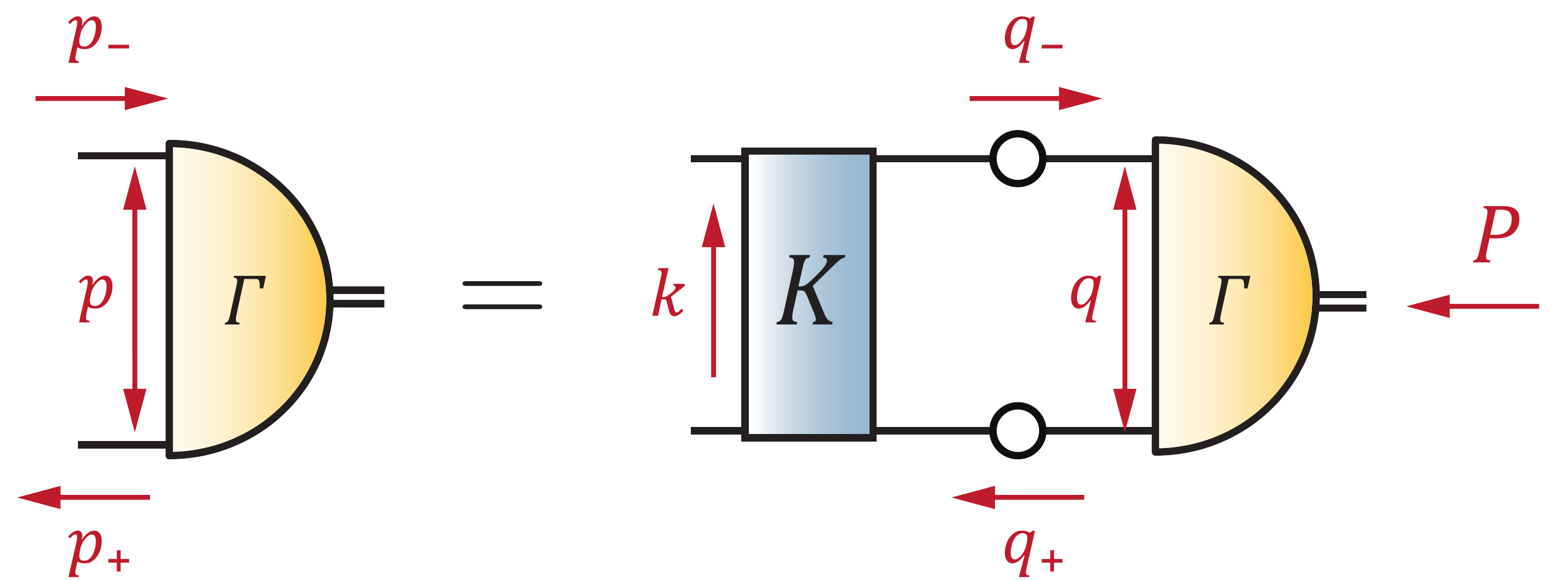} 
                    \caption{Left: Quark DSE from Eq.~\eqref{quark-dse}. The lines with circles denote the full quark propagator,
                             and the self-energy depends on the full gluon propagator and quark-gluon vertex.
                             Right: Bethe-Salpeter equation~\eqref{bse} for a meson.  }\label{fig:quark-dse}
            \end{figure*}

  The analogues of Eq.~\eqref{geom-series} in QCD are the DSEs,
  which are the exact, non-perturbative quantum equations of motion in a QFT.
  The quark DSE is shown in the left panel of Fig.~\ref{fig:quark-dse} and reads explicitly
  \begin{equation}\label{quark-dse}
     S(p)^{-1} = Z_2\,(i\slashed{p} + m_0) + \Sigma(p)\,, \qquad \Sigma(p) = - \frac{4 g^2}{3}\,Z_{\Gamma}\int \!\!\frac{d^4 q}{(2\pi)^4}\,
                                      i\gamma^\mu \, S(q) \, D^{\mu\nu}(k)\, \Gamma^\nu(l,k)\,.
  \end{equation}
  Here, $\Sigma(p)$ is the quark self-energy that  depends on the gluon propagator $D^{\mu\nu}(k)$ with momentum $k=q-p$, 
  the full quark-gluon vertex $g\,\Gamma^\mu(l,k)$ with relative momentum $l = (p+q)/2$ and total momentum $k$, and the tree-level vertex $g\,Z_\Gamma\,i\gamma^\mu$
  with renormalization constant $Z_\Gamma$.
  The prefactor $4/3$ comes from the color trace,
  $m_0$ is the bare current quark mass that enters in the QCD Lagrangian, and $Z_2$ is the quark renormalization constant.
    
    Note that the full quark propagator appears again inside the loop,
  so the DSE is an integral equation for $S(p)$. 
  It has the structural form $f(x)^{-1} = 1 - x$ from Eq.~\eqref{geom-series}, where $-x$ plays the role of the self-energy.
  Thus, if $g$ is small one can expand $S(p)$ in a series like in Eq.~\eqref{geom-series-3} by
  reinserting the equation at every instance where the quark propagator appears inside the loop, 
  doing the same for the gluon propagator and quark-gluon vertex, and stop after a few terms ---
  this reproduces the perturbative series for the quark propagator shown in Fig.~\ref{fig:quark-perturbative}.
  However, if the coupling is large we have no choice but to solve the equation directly.

  \subsection{An illustrative model}  
  
  At this stage you might be tempted to throw your hands in the air and 
  follow the standard textbook advice, which  portrays perturbation theory 
  as intuitive and easy to grasp while 
  nonperturbative phenomena are something more obscure and best left to specialists.
  However, the quark DSE serves as a hands-on example of how nonperturbative properties can be understood with surprising ease.
  There are simple analytically solvable models that capture the basic features in Fig.~\ref{fig:quark-mf}~\cite{Roberts:2007jh,Eichmann:2016yit}.
  An illustrative example is the Munczek-Nemirovsky model~\cite{Munczek:1983dx}, 
  where the quark-gluon vertex in Eq.~\eqref{quark-dse} is approximated by a bare vertex, $\Gamma^\nu(l,k) = i\gamma^\nu$,
  and the gluon propagator by a $\delta$ distribution, with a scale $\Lambda$ to ensure the correct dimension: 
  \begin{equation}\label{munczek-nemirovsky}
      \frac{4}{3}\,\frac{g^2}{(2\pi)^4}\,D^{\mu\nu}(k) \to \Lambda^2 \delta^4(k)\,\delta^{\mu\nu}\,.
  \end{equation}
  Shifting the integration momentum $d^4q \to d^4k$, the integral can be solved analytically,
  so there are no loop divergences and one can set the renormalization constants $Z_2 = Z_\Gamma = 1$. The DSE then turns into
  \begin{equation}
     S(p)^{-1} = i\slashed{p} + m_0 + \Lambda^2\,\gamma^\mu\,S(p)\,\gamma^\mu\,.
  \end{equation}
  Plugging in the decomposition~\eqref{quark-general} yields two algebraic equations for $A(p^2)$ and $M(p^2)$:
  \begin{equation}
     A(p^2) = 1 + \frac{2\Lambda^2}{A(p^2) \left( p^2 + M(p^2)^2\right)}\,, \qquad
     A(p^2) M(p^2) = m_0 + 2M(p^2)\,\frac{2\Lambda^2}{A(p^2) \left( p^2 + M(p^2)^2\right)}\,.
  \end{equation}
  In the chiral limit $m_0=0$, the chiral-symmetry preserving solution $M(p^2)=0$ is still possible 
  and yields a quadratic equation for $A(p^2)$, whose solution is
  \begin{equation}\label{mn-sol1}
     A(p^2) = \frac{1}{2} \left( 1 + \sqrt{1 + \frac{8\Lambda^2}{p^2}}\right).
  \end{equation}
  Here we picked the positive branch that reproduces the correct perturbative behavior $A(p^2)=1$. 
  However, there is also a chiral-symmetry breaking solution with $M(p^2) \neq 0$. Plugging the first equation above into the second yields
  \begin{equation}\label{mn-sol2}
     A(p^2) = 2\,, \qquad M(p^2) = \sqrt{\Lambda^2-p^2}\,.
  \end{equation}
  The two solutions are connected at $p^2=\Lambda^2$, as shown in the right panel of Fig.~\ref{fig:quark-mf}.
  When switching on the current-quark mass $m_0 \neq 0$, the solution becomes
  more cumbersome but the curves are now smooth.
  Note the analogy with the spontaneous magnetization of a magnet below a critical temperature 
  if $p^2$ is replaced by temperature, $M(p^2)$ by  magnetization and the current-quark mass $m_0$ by an external magnetic field.
  Loosely speaking, below a critical momentum $p^2 = \Lambda^2$, QCD dynamically generates a quark mass $M(p^2)$ even if the quarks in the QCD Lagrangian are massless.
  In more realistic DSE solutions, there is no sharp phase transition in the chiral limit and  one instead finds the behavior in the left panel of Fig.~\ref{fig:quark-mf}.

  How do we know the gluon propagator and quark-gluon vertex, which appear as inputs in the quark DSE?
  These quantities satisfy their own DSEs, which depend again on the quark propagator but also on higher $n$-point functions, so one arrives at a coupled system of equations
  which needs to be solved (Fig.~\ref{fig:dses}).
 Alternatively, if the gluon propagator and quark-gluon vertex are sufficiently well known,
  one can make a combined ansatz for them (an `effective interaction') and solve the quark DSE as a standalone equation.
  The second route has turned out especially useful in combination with hadron physics applications, to which we will turn in Sec.~\ref{sec:mesons}.
  
  \subsection{About renormalization}
  
  Before doing so, it is worthwile to briefly discuss the issue of renormalization, because the quark DSE illustrates the basic steps
  that are similar for other $n$-point functions. To this end, let us write the quark self-energy as $\Sigma(p) = i\slashed{p}\,\Sigma_A(p^2) + \Sigma_M(p^2)$,
  so that by Eq.~\eqref{quark-general} the quark DSE turns into two coupled scalar equations for $A(p^2)$ and $M(p^2)$:
  \begin{equation}\label{quark-dse-explicit}
       A(p^2) = Z_2 + \Sigma_A(p^2)\,, \qquad
       M(p^2) A(p^2) = Z_2 \,m_0 + \Sigma_M(p^2)\,.
  \end{equation}
  The self-energy integrals are logarithmically UV-divergent, so they must first be regularized, e.g., by integrating up to a finite cutoff. 
  In the renormalization step we set two boundary conditions for the integral equations by demanding
  \begin{equation}\label{ren-cond}
      A(\mu^2) \stackrel{!}{=} 1\,, \qquad
      M(\mu^2) \stackrel{!}{=} m\,.
  \end{equation}
  The second condition introduces the renormalized current-quark mass $m$ for a given quark flavor at some arbitrary renormalization scale \mbox{$p^2 = \mu^2$}.
  The first condition $A(\mu^2)=1$ is arbitrary and merely induces a  renormalization of $A(p^2)$;  if we chose another value
  this would not affect any observable.
  Eq.~\eqref{quark-dse-explicit} then fixes the bare parameters $Z_2$ and $m_0$, which are divergent if the cutoff is sent to infinity:
  \begin{equation}\label{z2}
      Z_2 = 1 - \Sigma_A(\mu^2)\,, \qquad
      m_0 = \frac{m - \Sigma_M(\mu^2)}{1 - \Sigma_A(\mu^2)}\,.
  \end{equation}
  Plugging this back  into Eq.~\eqref{quark-dse-explicit} yields the final form of the DSEs:
  \begin{equation}\label{quark-dse-explicit-2}
  \begin{split}
       A(p^2) = 1 + \Sigma_A(p^2) - \Sigma_A(\mu^2)\,,  \qquad 
       M(p^2) A(p^2) = m + \Sigma_M(p^2) - \Sigma_M(\mu^2)\,.
  \end{split}
  \end{equation}
  We see that the boundary conditions in Eq.~\eqref{ren-cond}  led to subtracted equations at the renormalization point;
  this is called a momentum-subtraction (MOM) scheme.
  Due to this subtraction, all divergences cancel and the  dressing functions $A(p^2)$ and $M(p^2)$ are finite.
  These equations can be solved iteratively: start with some guess for $A(p^2)$ and $M(p^2)$ (e.g., set them to 1),
  calculate $\Sigma_A(p^2)$ and $\Sigma_M(p^2)$ from Eq.~\eqref{quark-dse}, determine $\Sigma_A(\mu^2)$ and $\Sigma_M(\mu^2)$
  and correspondingly $Z_2$ from Eq.~\eqref{z2}, and arrive at new values for $A(p^2)$ and $M(p^2)$, which enter in the self-energy in the next step.
  The procedure is then repeated until convergence.

  In the renormalization process we traded the cutoff dependence for a dependence on the renormalization point $\mu$, so that $A(p^2)$ now also depends on $\mu$.
  The mass function $M(p^2)$, on the other hand, is renormalization-group invariant:
  After having solved the quark DSE, one can read off $M(p^2 = {\mu'}^2) = m'$ at another value $\mu'$, use this
  as the new renormalization condition instead of Eq.~\eqref{ren-cond} and solve the equations anew. 
  The resulting mass function is the same as before (so it does not depend on $\mu$),
  whereas $A(p^2)$ differs by a multiplicative constant. Note also that the renormalization scale $\mu$ is completely arbitrary;
  one may choose a very large value $\mu = 100$ GeV or just as well renormalize at $\mu = 0$.

\section{Mesons}\label{sec:mesons}

  The spontaneous breaking of chiral symmetry has important consequences for hadron physics.
  On the one hand, the large dynamical quark mass in Fig.~\ref{fig:quark-mf} 
  is the effective mass that enters in mesons and baryons as bound states of valence quarks.
  Without it, the spectrum of hadrons would look very different or they would not even be able to form.
  On the other hand, the spontaneous breaking of a global symmetry leads to massless Goldstone bosons in the spectrum.
  In QCD these are the pseudoscalar mesons, in particular the three pions whose masses of about 140 MeV are much smaller
  than the masses of other hadrons (which are of the order of 1 GeV).
  Therefore, the pion plays a special role in QCD: It is a bound state of a valence quark and antiquark 
  but at the same time the (pseudo-)Goldstone boson of spontaneous chiral symmetry breaking, whose mass would be exactly zero in the chiral limit.
  
  \subsection{Bethe-Salpeter equation}\label{sec:bse}
  
  These features are explicit in the Bethe-Salpeter equation (BSE) for the pion~\cite{Maris:1997hd}. 
  BSEs can be viewed as the quantum field-theoretical analogues of the Schrödinger equation in quantum mechanics:
  They are eigenvalue equations which allow one to calculate the spectrum and wave functions of hadrons.
  Here the word `analogue'  is important: 
  Because the Poincaré group is not compact (it contains Lorentz boosts),   its unitary representations must be infinite-dimensional.
  This is why quantum mechanics is not enough to describe special relativity and one should employ QFT, 
  which provides unitary representations on the infinite-dimensional Fock space and thus a probability interpretation. 
  The Bethe-Salpeter wave function, on the other hand,  is defined as
  \begin{equation}\label{bswf}
      [i\mathbf\Psi(p,P)]_{\alpha\beta} = \int d^4x\,e^{-ipx}\, \langle 0 \,| \,\mathsf{T}\,\psi_\alpha(x)\,\conjg\psi_\beta(0)\,|\, P, n \rangle \,,
  \end{equation}  
  where $P$ is the total onshell momentum of the hadron `$n$' defined by  quantum numbers like
  spin $J$, parity $P$, flavor, etc. Because the quark spinors transform under finite-dimensional (Dirac) representations
  of the Lorentz group, so does the Bethe-Salpeter wave function.
  As such, it does not have a direct probabilistic interpretation 
  like the wave functions in quantum mechanics. Nevertheless, it encodes the internal structure of a hadron
  and enters in observables through matrix elements like, e.g., form factors and parton distributions.

   The general form of the BSE for a meson made of a valence quark and antiquark is shown in the right panel of Fig.~\ref{fig:quark-dse} and reads
  \begin{equation}\label{bse}
     [\mathbf\Gamma(p,P)]_{\alpha\beta}  = \int \!\! \frac{d^4q}{(2\pi)^4} \, [\mathbf{K}(p,q,P)]_{\alpha\gamma,\delta\beta} \left[ S(q_+)\,\mathbf\Gamma(q,P)\,S(q_-)\right]_{\gamma\delta}\,.
  \end{equation}
   The Bethe-Salpeter amplitude $\mathbf\Gamma(p,P)$ is defined as the wave function 
     $\mathbf\Psi(p,P) = S(p_+)\,\mathbf\Gamma(q,P)\,S(p_-)$ without external quark propagator legs.
   The quark-antiquark kernel $\mathbf{K}(p,q,P)$ contains all possible two-particle irreducible interactions between quark and antiquark, i.e.,
   those that are not already generated by the iteration. $S(q_\pm)$ 
   is again the full quark propagator from Eq.~\eqref{quark-general}.
   There are further color and flavor indices stemming from the quark fields in Eq.~\eqref{bswf}, but we absorbed them in the Greek subscripts 
   which are therefore multi-indices in Dirac, color and flavor space. 
   
\begin{figure*}[t]
                    \centering
                    \includegraphics[width=0.5\textwidth]{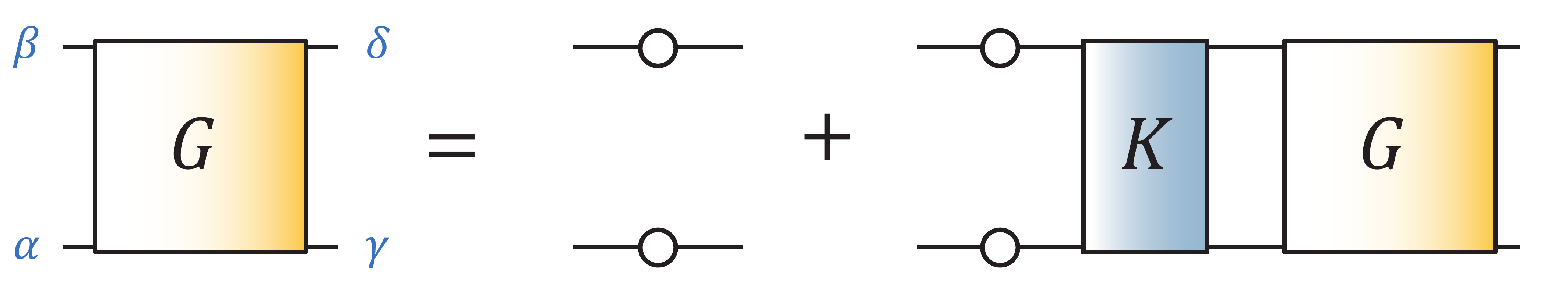}
                    \caption{Scattering equation for the quark-antiquark four-point function.}\label{fig:sc-eq}
            \end{figure*}

   How is the BSE derived? This is closely related 
   to the way how bound states appear in QFT in general. A QFT is fully specified by its (infinitely many) $n$-point correlation functions.
   Bound states appear as poles in these correlation functions.
   Take for example the quark-antiquark four-point function describing $q\bar{q}\to q\bar{q}$ scattering, which is shown in Fig.~\ref{fig:sc-eq}:
          \begin{equation}\label{four-point-function}
              \mathbf G_{\alpha\beta\gamma\delta}(x_1,x_2,x_3,x_4) = \langle 0 \,|\, \mathsf{T}\,\psi_\alpha(x_1)\,\conjg{\psi}_\beta(x_2)\,\psi_\gamma(x_3)\,\conjg{\psi}_\delta(x_4) \,|\, 0 \rangle\,.
          \end{equation}
          Inserting a complete set of states produces meson poles, 
          because a composite operator $\psi \conjg{\psi}$ can produce color singlet quantum numbers ($\mathbf{3}\otimes\bar{\mathbf{3}}=\mathbf{1}\oplus \mathbf{8}$). 
          At a given pole, the four-point function factorizes and defines the Bethe-Salpeter wave function as the residue at the pole:
          \begin{equation}\label{hadrons-poles}
             \mathbf G(p,q,P)  \stackrel{P^2 \,\to \, -m_n^2}{\longlonglongrightarrow} \frac{\mathbf\Psi(p,P)\,\conjg{\mathbf\Psi}(q,P)}{P^2 + m_n^2}\,   \,.
          \end{equation}
          The proof of this relation can be found, e.g., in Weinberg's book~\cite{Weinberg:1995mt}. In momentum space, the four-point function depends on three independent momenta
          (e.g., the relative momenta $p$, $q$ and the total momentum $P$), and it has poles whenever the total momentum goes onshell ($P^2 = -m_n^2$).
          Actually, Eq.~\eqref{hadrons-poles} can also be used to \textit{define} the wave function --- this is useful when the definition~\eqref{bswf}
          is no longer applicable, as in the case of unstable resonances where $|P,n\rangle$ is not an asymptotic state and the mass $m_n$ becomes complex.

          The fact that composite states appear as poles in $n$-point functions is a central feature of QFT: 
          This is what makes it possible to extract hadron properties experimentally,
          namely from bumps and peaks in experimental cross sections. It is also
          the underlying basis for
           theoretical treatments of 
          bound states in QFT such as e.g. lattice calculations,  effective field theories and functional methods.

           The actual derivation of the BSE in Eq.~\eqref{bse} goes as follows. 
           The four-point function $\mathbf{G}$ satisfies the scattering equation shown in Fig.~\ref{fig:sc-eq},
           which has the structural form
           \begin{equation}\label{sc-eq-G}
               \mathbf{G} = \mathbf{G}_0 + \mathbf{G}_0\,\mathbf{K}\,\mathbf{G}\,.
           \end{equation}
           Here, every multiplication stands for a four-dimensional momentum integration $\int d^4q_i /(2\pi)^4$,
           and $\mathbf{G}_0$ is the disconnected part of $\mathbf{G}$, i.e., the product of  quark and antiquark propagators.
           $\mathbf{K}$ is the $q\bar{q}$ irreducible kernel, which contains all interactions between quark and antiquark that do not fall apart
           by cutting two quark lines, i.e., those that are not already generated by the iteration
           $\mathbf{G} = \mathbf{G}_0 + \mathbf{G}_0\,\mathbf{K}\,\mathbf{G}_0 + \mathbf{G}_0\,\mathbf{K}\,\mathbf{G}_0\,\mathbf{K}\,\mathbf{G}_0 + \dots$.
           Now, at a given meson pole location the four-point function $\mathbf{G}$ satisfies Eq.~\eqref{hadrons-poles}.
           Comparing residues on both sides of the scattering equation, and because $\mathbf{G}_0$ does not have meson poles,
           we arrive at the homogeneous BSE $\mathbf\Psi = \mathbf{G}_0\,\mathbf{K}\,\mathbf{\Psi}$ 
           which holds at the pole.
           With $\mathbf{\Psi} = \mathbf{G}_0\,\mathbf{\Gamma}$, this can also be expressed in terms of the Bethe-Salpeter amplitude
           as in Eq.~\eqref{bse}, $\mathbf\Gamma = \mathbf{K}\,\mathbf{G}_0\,\mathbf{\Gamma}$. 
           
\begin{figure*}[t]
                    \centering
                    \includegraphics[width=0.8\textwidth]{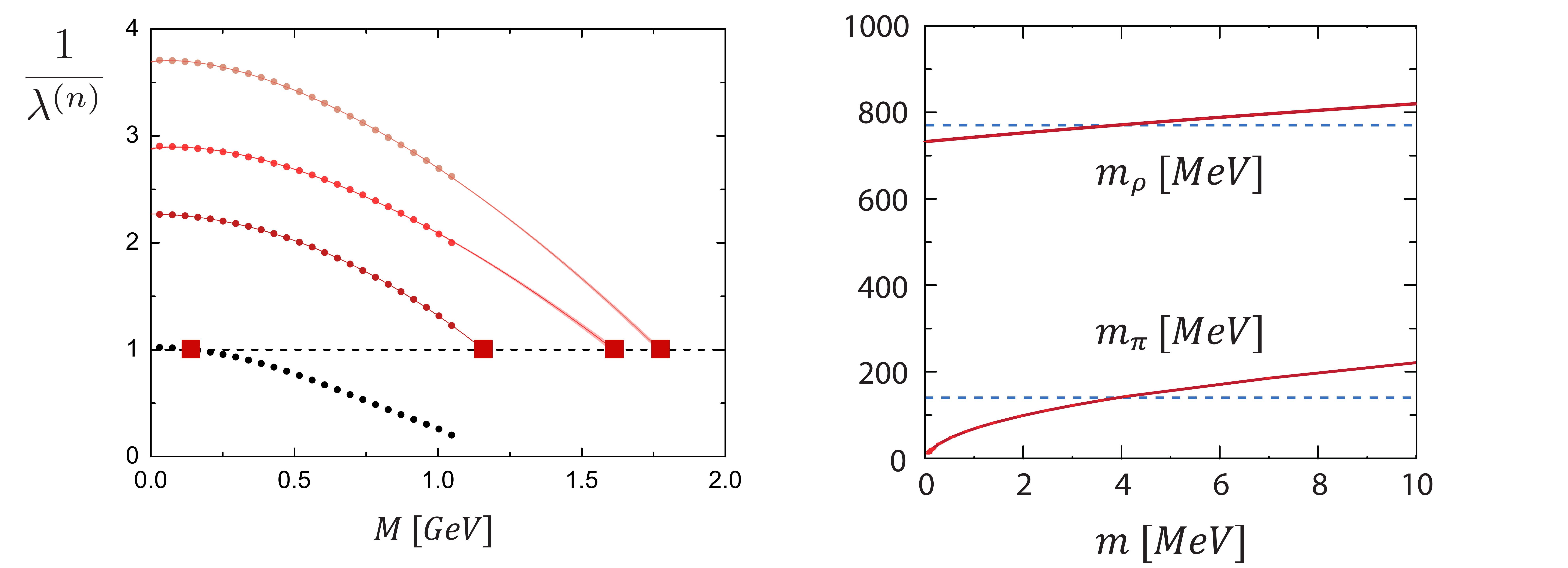}
                    \caption{Left: inverse  eigenvalues $\lambda^{(n)}(P^2 = -M^2)$ of the pion BSE as functions of $M$~\cite{Eichmann:2016nsu}.
                             The points are numerical results and the bands are fits.
                             Whenever an eigenvalue crosses the line 1, one has found a physical solution $M=m_n$ (red squares).
                             The calculation stops at $M \approx 1.1$ GeV because for larger values one would need to circumvent
                             quark singularities in the integrand. Right: pion and $\rho$-meson masses obtained from their Bethe-Salpeter equations
                             and plotted over the current-quark mass $m$ (which is an input in the quark DSE, Eq.~\eqref{ren-cond}).
                             The horizontal dashed  lines are the experimental values~\cite{Eichmann:2016yit}. The pion mass satisfies the GMOR relation~\eqref{gmor}. 
                             }\label{fig:evs}
            \end{figure*}

           A few  remarks are in order:
       \begin{itemize}
       
       \item It is  useful to define the scattering matrix $\mathbf{T}$ as the connected part of  
           $\mathbf{G} = \mathbf{G}_0 + \mathbf{G}_0\,\mathbf{T}\,\mathbf{G}_0$, so that Eq.~\eqref{sc-eq-G} turns into the scattering equation 
           $\mathbf{T} = \mathbf{K} + \mathbf{K}\,\mathbf{G}_0\,\mathbf{T}$.
           Because any pole in $\mathbf{G}$ must also show up in $\mathbf{T}$,  Eq.~\eqref{hadrons-poles}  holds for $\mathbf{T}$  alike if $\mathbf{\Psi}$ 
           is replaced by $\mathbf{\Gamma}$. This yields again the  homogeneous BSE $\mathbf\Gamma = \mathbf{K}\,\mathbf{G}_0\,\mathbf{\Gamma}$.

   \item  The homogeneous BSE is an eigenvalue equation. If we label the ground state in a given $J^{PC}$ channel by $n=0$ and the excited states by $n \geq 1$, we have
   \begin{equation}
       \mathbf{K}\,\mathbf{G}_0\,\mathbf{\Gamma}^{(n)} = \lambda^{(n)}\,\mathbf\Gamma^{(n)}\,.
   \end{equation}
   That is, if a given eigenvalue $\lambda^{(n)}$ satisfies $\lambda^{(n)} = 1$, we have found a pole in $\mathbf{G}$ and $\mathbf{T}$.
   This is also obvious when we write the  scattering equations in their inverse forms,
   $\mathbf{G}^{-1} = \mathbf{G}_0^{-1} - \mathbf{K}$ and $\mathbf{T}^{-1} = \mathbf{K}^{-1} - \mathbf{G}_0$. 
   A pole in $\mathbf{G}$ and $\mathbf{T}$ implies a zero in $\mathbf{G}^{-1}$ and $\mathbf{T}^{-1}$, which in turn requires
   an eigenvalue for $\mathbf{K}\,\mathbf{G}_0$ which is 1.

   \item  In practice, it is useful to work out the tensor decomposition of the amplitude $\mathbf\Gamma(p,P)$ in order to obtain
    Lorentz-invariant equations for its dressing functions.
  The pion amplitude has a Dirac, flavor and color part:
    \begin{equation}\label{pion-amp}
       \mathbf\Gamma(p,P) = \left( f_1 + f_2\,i\slashed{P} + f_3\,(p\cdot P)\,i\slashed{p} + f_4\,[\slashed{p},\slashed{P}]\right) \gamma_5 \otimes \text{Flavor} \otimes \text{Color} \,.
    \end{equation}
    The first part is a matrix in Dirac space which depends on four linearly independent Dirac tensors.
    We denoted their  dressing functions by $f_i(p^2,z,P^2=-m_\pi^2)$, where $z = \hat{p}\cdot\hat{P}$ is the angular variable 
    and a hat denotes a normalized unit vector, $\hat{p}^\mu = p^\mu / \sqrt{p^2}$. 
    After taking Dirac, color and flavor traces, the BSE~\eqref{bse} assumes the Lorentz-invariant form
    \begin{equation}
       \lambda^{(n)}(P^2)\,f_i^{(n)}(p^2,z,P^2) = \int_0^\infty dq^2 \int_{-1}^1 dz'\,K_{ij}(p^2,q^2,z,z',P^2)\,f_j^{(n)}(q^2,z',P^2)\,.
    \end{equation}
    If we further discretize the variables $p^2$ and $z$, this becomes a matrix-vector eigenvalue equation whose outcomes are the eigenvalues $\lambda^{(n)}(P^2)$
    and corresponding eigenvectors. As shown in Fig.~\ref{fig:evs}, if an eigenvalue crosses the line 1, this implies a pole in the scattering matrix
    and thus a physical solution.

   \item   The homogeneous BSE holds for bound states and resonances alike. 
   The pole condition~\eqref{hadrons-poles} does not care if the mass $m_n$ is real or complex.
   For a bound state, the condition $\lambda^{(n)}(P^2 = -m_n^2)=1$ has a solution for real $m_n$,
   while for a resonance the solution lies in the complex plane on a higher Riemann sheet. 
   Then,  by solving the BSE for complex values of $P^2$ on the first sheet, the resulting eigenvalues $\lambda^{(n)}(P^2)$ 
   can be analytically continued to the second sheet to solve the pole condition, see e.g.~\cite{Santowsky:2020pwd} for examples.
   In practice, solving for complex $P^2$ can be technically challenging and may require contour deformations.

   \end{itemize}

  \subsection{Chiral properties of the pion}
  
  What does the kernel $\mathbf{K}$ look like? So far we only know that it must be two-particle irreducible, because this is how the scattering equation~\eqref{sc-eq-G} was defined.
  Clearly, somehow $\mathbf{K}$ must be    aware of chiral symmetry to produce a massless pion in the chiral limit.
  To this end, let us make a brief detour and sketch the chiral properties of the pion (see~\cite{Maris:1997hd,Eichmann:2016yit} for details).
  Consider the axialvector and pseudoscalar currents,
  whose matrix elements define the pion's electroweak decay constant $f_\pi$ and its `pseudoscalar decay constant' $r_\pi$:
  \begin{equation}\label{fpi-rpi}
  \begin{array}{rl}
     j_5^\mu(x) &\!\!\!\!= Z_2\,\conjg{\psi}(x)\,i \gamma^\mu \gamma_5 \,\psi(x)\,, \\[1mm]
     j_5(x) &\!\!\!\!= Z_2 Z_m\,\conjg{\psi}(x)\,i\gamma_5 \,\psi(x)\,,
  \end{array} \quad
  \begin{array}{rl}
     \langle 0 \,|\, j_5^\mu(x) \,|\, P \rangle &\!\!\!\!= \langle 0 \,|\, j_5^\mu(0) \,|\, P \rangle \,e^{ix\cdot P} = i f_\pi\, P^\mu e^{ix\cdot P}\,, \\[1mm]
     \langle 0 \,|\, j_5(x) \,|\, P \rangle &\!\!\!\!= \langle 0 \,|\, j_5(0) \,|\, P \rangle \,e^{ix\cdot P} = r_\pi\,e^{ix\cdot P} \,.
  \end{array}
  \end{equation}
  Here we denoted the pion state by $|P\rangle$, $Z_2$ is the quark renormalization constant, and $Z_m$ is the quark mass renormalization constant (the bare and renormalized masses are related by $m_0 = Z_m\,m$).
  One can see from Eq.~\eqref{bswf} that these are just the four-momentum integrals over the Bethe-Salpeter wave function $\mathbf\Psi(p,P)$,
  once taking the trace with $ \gamma^\mu \gamma_5$ and once with $\gamma_5$:  
  \begin{equation}\label{fpi-rpi2}
      f_\pi \,P^\mu  =  Z_2 \int \!\! \frac{d^4p}{(2\pi)^4}\,\text{Tr}\left\{ i\gamma_5 \gamma^\mu \mathbf\Psi(p,P) \right\} \,, \qquad
      r_\pi =   Z_2 Z_m\int \!\! \frac{d^4p}{(2\pi)^4}\,\text{Tr}\left\{ \gamma_5  \mathbf\Psi(p,P) \right\}\,.
  \end{equation}  
  Once the pion wave function is known from its BSE, it is thus an easy task to calculate $f_\pi$.
  
  Now consider the PCAC relation $\partial_\mu j_5^\mu = 2m j_5$, which follows from the Noether theorem and reflects
  the explicit breaking of the axialvector symmetry in Eq.~\eqref{va-tf} away from the chiral limit.
  Plugged into Eq.~\eqref{fpi-rpi}, we immediately find the relation
  \begin{equation}\label{pcac}
     f_\pi\,m_\pi^2 = 2m r_\pi \,,
  \end{equation}
  which is valid for all current-quark masses. In particular, this implies $f_\pi\,m_\pi^2 = 0$ in the chiral limit. 
  So, if one can show that the pion decay constant $f_\pi$ does \textit{not} vanish in the chiral limit, then $m_\pi$ must be zero.
  
  The right place to look for such a relation is  the axialvector Ward-Takahashi identity (axWTI), which is the PCAC relation evaluated for the three-point functions
  $\langle 0 \, | \, \mathsf{T}\,\psi_\alpha(x)\,\conjg \psi_\beta(y) \,j_5^\mu(z) \, | \, 0 \rangle$ and  
  $\langle 0 \, | \, \mathsf{T}\,\psi_\alpha(x)\,\conjg \psi_\beta(y) \,j_5(z) \, | \, 0 \rangle$.
  These quantities  have pion poles, whose residues are  proportional to the pion amplitude $\mathbf\Gamma(p,P)$.
  However, in the axWTI these poles cancel each other and one finds the chiral-limit relation
  \begin{equation}\label{B/fpi}
      f_\pi\,\mathbf\Gamma(p,0) = A(p^2)\,M(p^2)\,\gamma_5\,,
  \end{equation}
  where $A(p^2)$ and $M(p^2)$ are the quark dressing functions from Sec.~\ref{sec:quarks}.
  If chiral symmetry is dynamically broken and hence $M(p^2) \neq 0$, then the r.h.s. of this equation is nonzero, so also $f_\pi$
  cannot vanish in the chiral limit.  
   Eq.~\eqref{pcac} then entails  $m_\pi = 0$ in the chiral limit. In other words, we just proved the Goldstone theorem!
  Note  that by Eq.~\eqref{B/fpi} the pion amplitude in the chiral limit is already determined by the quark propagator,
  so in the chiral limit the pion BSE and quark DSE are equivalent.
  
  Finally, if we take the  trace of Eq.~\eqref{B/fpi} with $S(p)\,\gamma_5\,S(p)$ and compare with~\eqref{fpi-rpi2}, we obtain
  $f_\pi\,r_\pi = - \langle \conjg{\psi}\psi \rangle/N_f$ in the chiral limit, where the quark condensate is the trace
  over the quark propagator:
  \begin{equation}\label{quark-condensate}
     - \frac{\langle \conjg{\psi}\psi \rangle}{N_f} = Z_2 Z_m\,N_c\int \!\!\frac{d^4p}{(2\pi)^4} \,\text{Tr}\,S(p) = Z_2 Z_m\,\frac{N_c}{(2\pi)^2}\int dp^2\,\frac{p^2}{A(p^2)}\,\frac{M(p^2)}{p^2 + M(p^2)^2}\,.
  \end{equation}
  Therefore, also the quark condensate is directly related to the quark mass function.
  Plugged into Eq.~\eqref{pcac}, this returns the Gell-Mann-Oakes-Renner (GMOR) relation, which says   
  that near the chiral limit the squared pion mass is proportional to the current-quark mass:
  \begin{equation}\label{gmor}
     f_\pi^2 m_\pi^2 = -2m \,\langle  \conjg{\psi}\psi \rangle / N_f\,.
  \end{equation}
  
  \subsection{Bethe-Salpeter kernel}  
  
  The relations so far are exact. This implies  that the pion obtained from the BSE is automatically massless in the chiral limit  
  if (a)  the BSE kernel satisfies chiral symmetry through the axWTI,
  and (b) chiral symmetry is dynamically broken such that $M(p^2) \neq 0$ in the chiral limit.
  As we saw in Sec.~\ref{sec:quarks}, the condition (b) is automatic in the DSE. Concerning (a),
  the axWTI can be rewritten as  a consistency relation between the BSE kernel and the quark propagator, which must be satisfied
  by any practical truncation. 
  
  In principle one can derive the kernel  from the quantum effective action, which leads to a systematic expansion
  in powers of loops and powers of $n$-point functions.
  The leading term in such an expansion is the  rainbow-ladder truncation,
  which amounts to a gluon exchange between quark and antiquark.
  Here  the quark-gluon vertex
  is reduced to its leading tensor $i\gamma^\mu$, and its dressing functions is combined with that of the gluon propagator 
  into an effective interaction $\alpha(k^2)$  which depends on the gluon momentum $k$ only:
  \begin{equation}\label{rl-kernel}
     \mathbf{K}_{\alpha\gamma,\delta\beta}(p,q,P) = Z_2^2\,\frac{4\pi\alpha(k^2)}{k^2} \,(\mathsf{t}_a)_{AC}\,(\mathsf{t}_a)_{DB}\,(i\gamma^\mu)_{\alpha\gamma}\,T^{\mu\nu}_k (i\gamma^\nu)_{\delta\beta}\,.
  \end{equation}  
  This automatically satisfies the axWTI and thus the chiral symmetry constraints.
  $Z_2$ is the quark renormalization constant, $k^\mu$ the gluon momentum, $T^{\mu\nu}_k = \delta^{\mu\nu} - k^\mu k^\nu/k^2$ the transverse projector for Landau gauge, and $\mathsf{t}_a = \lambda_a/2$ are the generators of $SU(3)_c$ in the fundamental representation (the $\lambda_a$ are the eight Gell-Mann matrices).
  
  A popular ansatz for the effective interaction $\alpha(k^2)$ is the 
  Maris-Tandy model~\cite{Maris:1999nt}, $\alpha(k^2) = \pi\,\eta^7 x^2\,e^{-\eta^2 x} + \alpha_\text{UV}(x)$ with $x=k^2/\Lambda^2$.
  Here the  second term $\alpha_\text{UV}(x)$ ensures the correct perturbative behavior at large momenta but is otherwise not  important.
  The relevant term is the first one, which effectively depends on only one parameter, namely the scale $\Lambda = 0.72$ GeV that is chosen to reproduce the 
  experimental pion decay constant $f_\pi \approx 92$ MeV. If one varies the shape parameter $\eta$ within a certain range, 
  the shape of the interaction may look very different, but many observables calculated with it are insensitive to this change.
  For this reason, also other forms of $\alpha(k^2)$ proposed in the literature lead to very similar results for observables~\cite{Qin:2011dd}.

   With this one has all the tools ready to do spectroscopy calculations. The right panel in Fig.~\ref{fig:evs} shows the pion and $\rho$ mass
   obtained from their respective BSEs as a function of the current-quark mass, which is an input to the quark DSE, Eq.~\eqref{ren-cond}.
   One can see that the GMOR relation~\eqref{gmor} for the pion mass holds, $m_\pi \sim \sqrt{m}$. The pion mass vanishes in the chiral limit,
   while the $\rho$ mass does not. Both features are direct consequences of dynamical chiral symmetry breaking and 
   the associated dynamical mass generation.
   
   The DSE/BSE approach using rainbow-ladder  has been used for many practical calculations: masses and decay constants of light, heavy and heavy-light mesons;
   electromagnetic and transition form factors of mesons; parton distribution functions for pions and kaons; 
   scattering amplitudes like $\pi\pi\to\pi\pi$, $\gamma\pi\to\pi\pi$, etc; see e.g.~\cite{Maris:2005tt,Bashir:2012fs,Eichmann:2016yit} and references therein.
   Also three- and four-quark states have been successfully calculated in this approach, as we will see in Sections~\ref{sec:baryons} and~\ref{sec:tetraquarks}.
   As stated above, the only relevant parameters are the scale $\Lambda$, which is fixed by $f_\pi$, and the current-quark masses which are fixed by
   certain meson masses --- like in Fig.~\ref{fig:evs}, where one can set the light $u/d$ mass by the experimental pion mass (horizontal dashed line).
   In many cases this leads to predictions that lie within $\lesssim 5 \dots 10\%$ of experimental results.
   Noteable exceptions are scalar and axialvector mesons (we will say more about scalar mesons in Sec.~\ref{sec:tetraquarks}), 
   some properties of heavy-light mesons (which can be improved~\cite{Gao:2024gdj} by feeding information on the quark propagator back into $\alpha(k^2)$),
   and the radial excitations in various $J^{PC}$ channels.
   Obviously, rainbow-ladder is not full QCD, but it is a very useful starting point for a wide array of applications.
   It is also the turning point between QCD modelling and a systematic expansion in QCD's $n$-point functions,
   for which one must solve coupled systems of DSEs. 
    Such investigations beyond rainbow-ladder are  ongoing, see e.g.~\cite{Chang:2009zb,Eichmann:2016yit,Huber:2020keu,Eichmann:2021zuv,Ferreira:2023fva}.

  \subsection{An illustrative model (continued)} 

To conclude this section and illustrate some features in the chiral limit, let us again employ the  toy model from Eq.~\eqref{mn-sol2}.
Plugging it into the quark condensate~\eqref{quark-condensate}, with $M(p^2) = 0$ for $p^2 > \Lambda^2$, yields for a single flavor (e.g., an up quark)
\begin{equation}
        - \langle \conjg{u}u \rangle = \frac{N_c}{(2\pi)^2}\int_0^{\Lambda^2} dp^2\,p^2\,\frac{\sqrt{\Lambda^2-p^2}}{2\Lambda^2}
                                     = \frac{2}{15}\,\frac{N_c}{(2\pi)^2} \,\Lambda^3   \,.
\end{equation}
Setting $\Lambda = 1$ GeV returns a reasonable value for the condensate: $- \langle \conjg{u}u \rangle \approx (220\,\text{MeV})^3$.
Concerning the BSE, in the chiral limit we can drop the dependence on the total momentum because the onshell pion satisfies $P^2 = 0$ and thus $P^\mu = 0$, 
so the amplitude has only one tensor $\mathbf\Gamma(p) = f_1(p^2)\,\gamma_5$. 
Using the rainbow-ladder truncation~\eqref{rl-kernel} with the $\delta$-distribution~\eqref{munczek-nemirovsky}, the BSE at $P^2=0$  turns into
\begin{equation}
   \mathbf\Gamma(p) = \frac{4 g^2}{3} \int \!\!\frac{d^4q}{(2\pi)^4} \,i\gamma^\mu S(q)\,\mathbf\Gamma(q)\,S(q)\,i\gamma^\nu\,D^{\mu\nu}(k)  
    \; \to \;  - \Lambda^2\,\gamma^\mu S(p)\,\mathbf\Gamma(p)\,S(p)\,.
\end{equation}
If the amplitude is non-zero, $f_1(p^2)$ drops out from both sides of the equation and we obtain
\begin{equation}
  1 = \Lambda^2\,\gamma^\mu \gamma_5\,S(p)\,\gamma_5\,S(p)\gamma^\mu = \Lambda^2 \gamma^\mu S(-p)\,S(p)\,\gamma^\mu = \frac{4\Lambda^2}{A(p^2)^2 \left( p^2 + M(p^2)^2\right)}\,,
\end{equation}
which is satisfied by Eq.~\eqref{mn-sol2}.
In other words, the BSE has eigenvalue 1 at $P^2 = 0$.
As advertised, the pion obtained from the BSE in the chiral limit is automatically massless because the kernel respects chiral symmetry.
Although in this simple model the BSE does not tell us what $f_1(p^2)$ itself looks like, we already know from Eq.~\eqref{B/fpi} that
$f_1(p^2) = A(p^2) M(p^2)/f_\pi$ must hold generally.
This can be plugged back into Eq.~\eqref{fpi-rpi2} to get an expression for $f_\pi$ in the chiral limit, which
is known as the Pagels-Stokar formula~\cite{Pagels:1979hd}:
\begin{equation}
   f_\pi^2 = \frac{N_c\,Z_2}{4\pi^2} \int dp^2\,p^2\,\frac{M(p^2)}{A(p^2)}\,\frac{1}{\left( p^2 + M(p^2)^2 \right)^2} \left( M(p^2) - \frac{p^2}{4}\,\frac{dM(p^2)}{dp^2}\right)\,.
\end{equation}
The formula is not exact because $f_2$, $f_3$ and $f_4$ in Eq.~\eqref{pion-amp} do not vanish in the chiral limit, only $P^\mu$ does;
but since they are all (strongly) suppressed compared to $f_1$ it is a very good approximation.
In the Munczek-Nemirovsky model, we obtain
\begin{equation}
   f_\pi = \sqrt{\frac{N_c}{2}}\,\frac{\Lambda}{4\pi}\,.
\end{equation}
With $\Lambda = 1$ GeV this gives $f_\pi = 97$ MeV, which is again in the right ballpark. 
Finally, combining the results for $f_\pi$ and $\langle u \bar{u}\rangle$, the GMOR relation~\eqref{gmor} turns into $m_\pi^2 = \frac{32}{15}\,\Lambda m \approx 2\Lambda m$.

\newpage

            \begin{figure*}[t]
                    \centering
                    \includegraphics[width=0.8\textwidth]{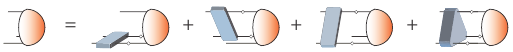}
                    \caption{Three-quark Faddeev equation with two- and three-body irreducible kernels.}\label{fig:faddeev}
            \end{figure*}

\section{Baryons}\label{sec:baryons}

 The proton is the only truly stable hadron. As such, it is a central ingredient in hadron structure
 experiments: from elastic and deep inelastic $ep$ scattering to $pp$ and $p\bar{p}$ reactions, $N\pi$ scattering, pion photo- and
 electroproduction, nucleon Compton scattering and more. Searches for physics beyond the Standard
 Model are usually also performed on protons and nuclei, which  requires a precise understanding of the hadron physics background.
  The proton has a mass of almost 1 GeV, but the current quarks only contribute 2--5 MeV each.
  We argued that dynamical mass generation can explain such discrepancies, as exemplified in Fig.~\ref{fig:evs} for the $\rho$ meson.
  Does the same apply for the proton?
  
  The study of baryons also does not stop at the proton. The nucleon and $\Delta$ excitation spectra
  have been subject to many investigations in quark models, and new photo- and electroproduction experiments
  have added several new states to the PDG~\cite{Klempt:2009pi,Tiator:2011pw,Aznauryan:2011qj,Crede:2013kia,Gross:2022hyw,Thiel:2022xtb}. 
  Especially interesting are also hyperons, i.e., baryons with strange quarks. 
  Compared to  light baryons, the experimentally known hyperon spectrum
  is still rather sparse  even though it houses many interesting questions. Hyperons are important for hypernuclei
  and neutron stars, and the flavor-singlet $\Lambda(1405)$ made of $uds$ is the prime candidate for a five-quark
  admixture in the light baryon sector~\cite{Mai:2020ltx}. In recent years, much experimental progress has also been made for 
  charmed baryons where several new states have been found at LHCb~\cite{LHCb-FIGURE-2021-001-report}.

  \subsection{Baryons made of three valence quarks} 

\begin{wrapfigure}[35]{r}{0.3\textwidth}
\vspace{-5mm}
\includegraphics[width=\linewidth]{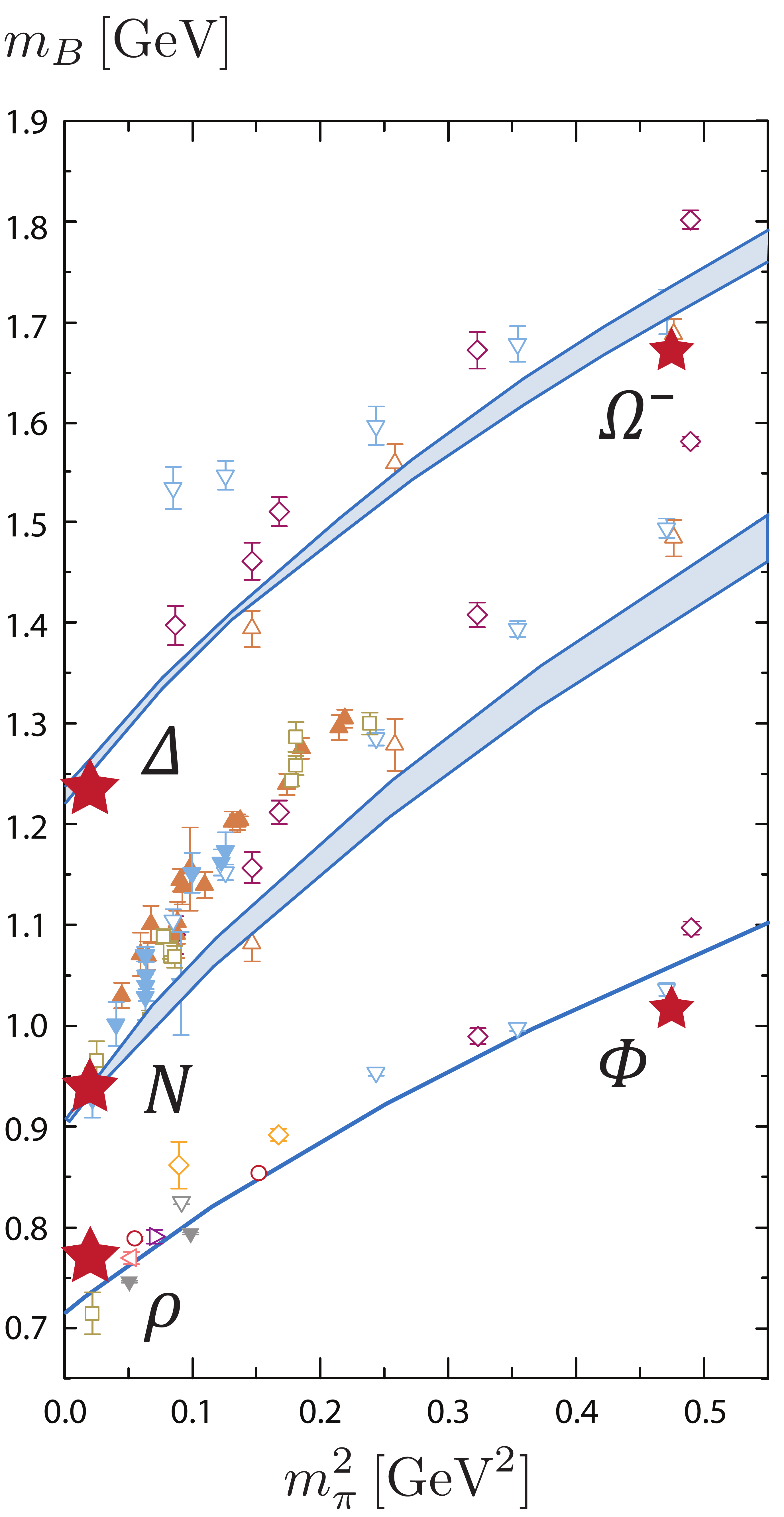} 
\caption{Nucleon, $\Delta$-baryon and $\rho$-meson masses as a function of the current-quark mass. The bands are the calculated masses 
and show the sensitivity to the shape parameter $\eta$. The stars are the experimental masses, and the points are lattice results for comparison~\cite{Eichmann:2016yit}.}
\label{fig:3body}
\end{wrapfigure}
  
  The three-body BSE is the covariant Faddeev equation shown in Fig.~\ref{fig:faddeev}.
  In this case, the kernel is a sum of permuted two-quark irreducible kernels plus a  three-quark irreducible kernel.
  It turns out that the leading three-body force, which is the three-gluon vertex connecting  all three quark legs,
   vanishes by the color trace. This is simply due to the relation
  \begin{equation}
     f_{abc}\,(\mathsf{t}_a)_{FM} \,(\mathsf{t}_b)_{GN} \,(\mathsf{t}_c)_{HR} \,\varepsilon_{MNR} = 0\,,
  \end{equation}
  where $f_{abc}$ is the color factor of the three-gluon vertex, the $SU(3)$ generators $\mathsf{t}_a$ belong to the  quark-gluon vertices,
  and $\varepsilon_{MNR}$ is the color wave function of the baryon. Of course, this does not prevent the appearance
  of higher $n$-point functions in the three-body kernel, like  quark-two-gluon or four-gluon vertices, but the simplest option is discarded. 
 This seems to point  towards two-body correlations as the dominant forces in baryons. Neglecting the three-body kernel,
  the Faddeev equation can then be solved using the very same setup as in the meson sector, in particular with the same two-body kernel.

  The main complication  arises from the structure of the wave functions, which are much richer compared to the case of mesons in Eq.~\eqref{pion-amp}.
  For a baryon with spin $J=1/2$, the Faddeev amplitude reads
    \begin{equation}\label{baryon-amp}
       \mathbf\Gamma(p,q,P)_{\alpha\beta\gamma\delta} = \sum_{i=1}^{64} f_i(p^2,q^2,p\cdot q, p\cdot P, q\cdot P, P^2) \, \tau_i(p,q,P)_{\alpha\beta\gamma\delta}  \otimes \text{Flavor} \otimes \varepsilon_{ABC} \,.
    \end{equation}
  It has four Dirac indices, three for the valence quarks and one for the nucleon. This leads to 64 Dirac tensors
  for baryons with spin $J=1/2$ and 128 tensors for $J=3/2$. Furthermore, the amplitude is a four-point function, so it depends
  on three independent momenta, e.g., two relative momenta $p$, $q$ and the total onshell momentum $P$.
  From three momenta one can form six Lorentz invariants (with $P^2 = -m_B^2$, where $m_B$ is the mass of the baryon), so the 64 dressing functions depend on six variables.

  Despite this, the Faddeev equation can be solved without any approximations on the amplitudes, i.e.,
  including all tensors and Lorentz invariants in the system~\cite{Eichmann:2009qa,Eichmann:2011vu}. The resulting nucleon and $\Delta$-baryon masses
  are shown in Fig.~\ref{fig:3body}. Here the rainbow-ladder truncation~\eqref{rl-kernel} was employed, with the same  
  Maris-Tandy interaction $\alpha(k^2)$ which depends on one relevant parameter $\Lambda$ fixed by $f_\pi$.
  Interestingly,  the resulting nucleon mass is $940$ MeV --- it appears that rainbow-ladder is a \textit{very} good starting point 
  towards understanding not only the properties of mesons but also those of baryons.
  Similar calculations have also been successfully employed for other octet and decuplet baryons as well as for heavy baryons~\cite{Sanchis-Alepuz:2014sca,Qin:2018dqp,Qin:2019hgk}.

  \newpage

            \begin{figure*}[t]
                    \centering
                    \includegraphics[width=0.8\textwidth]{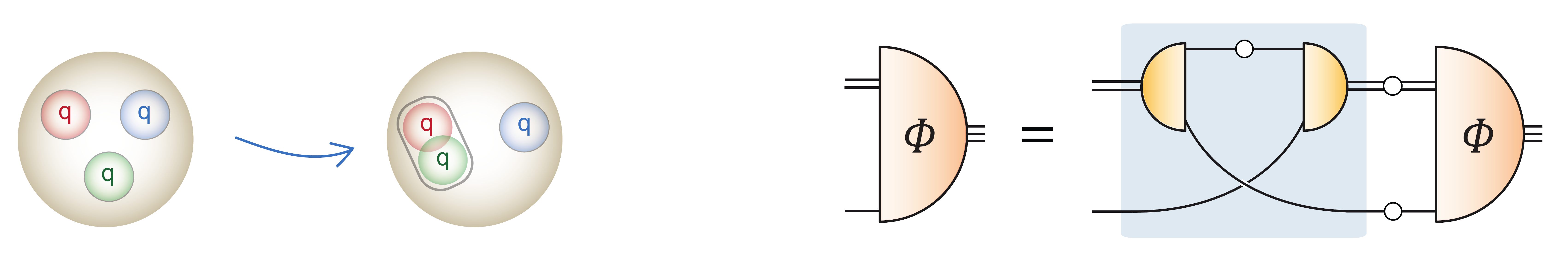}
                    \caption{Left: Two quarks inside a baryon can cluster to a diquark. Right: Quark-diquark BSE.}\label{fig:qdq}
            \end{figure*}
  
  \subsection{Diquarks} 
  
  What is the internal structure of baryons? A longstanding idea is that two quarks inside a baryon may cluster to a colored diquark~\cite{Anselmino:1992vg,Barabanov:2020jvn}.
  From 
  \begin{equation}\label{color-baryon}
      \mathbf{3}\otimes \mathbf{3} \otimes \mathbf{3} = ( \overline{\mathbf{3}} \oplus \mathbf{6} ) \otimes \mathbf{3} = 
      \mathbf{1} \oplus \mathbf{8} \oplus \mathbf{8} \oplus \mathbf{10}
  \end{equation}
  one can see that quarks may form color-antitriplet ($\overline{\mathbf{3}}$) and color-sextet ($\mathbf{6}$) diquarks, but 
  only a $\overline{\mathbf{3}}$ diquark can combine with the remaining quark to form a color-singlet baryon.
  From the color algebra one can also show that the quark-quark interaction in the $\overline{\mathbf{3}}$ channel
  is attractive and half as strong as that for color-singlet mesons.

  Traditionally, one argument in favor of  diquark correlations in baryons has been the so-called `missing-resonances problem',
  which is the observation that nonrelativistic quark models predict too many states in the baryon excitation spectrum.
  A three-quark system has two excitation modes, a `$\rho$ oscillator' between two quarks and a `$\lambda$ oscillator'
  between those  quarks and the remaining quark. If quarks formed pointlike diquarks, 
  this would freeze the $\rho$ mode and thus reduce the number of excitations.
  However, over the last decade new states have been found in photo- and electroproduction,
  which calls this simple idea into question~\cite{Nikonov:2007br}; moreover,
  static diquarks that cannot exchange their roles would also break the permutation symmetry.
  On the other hand, non-pointlike diquark correlations in the form of two-body clusters appear quite naturally in the three-body equation.
  With only a few assumptions~\cite{Eichmann:2016yit}, the Faddeev equation in Fig.~\ref{fig:faddeev} can be converted into a  quark-diquark BSE which is shown in Fig.~\ref{fig:qdq}.
  Here the interaction proceeds through a ping-pong quark exchange between quark and diquark, 
  so that the diquarks permanently exchange their roles. Furthermore, this two-body BSE can be solved using the very same
  ingredients as for the three-body equation, which allows for a direct comparison between the two approaches.

\begin{figure*}[t]
                    \centering
                    \includegraphics[width=0.9\textwidth]{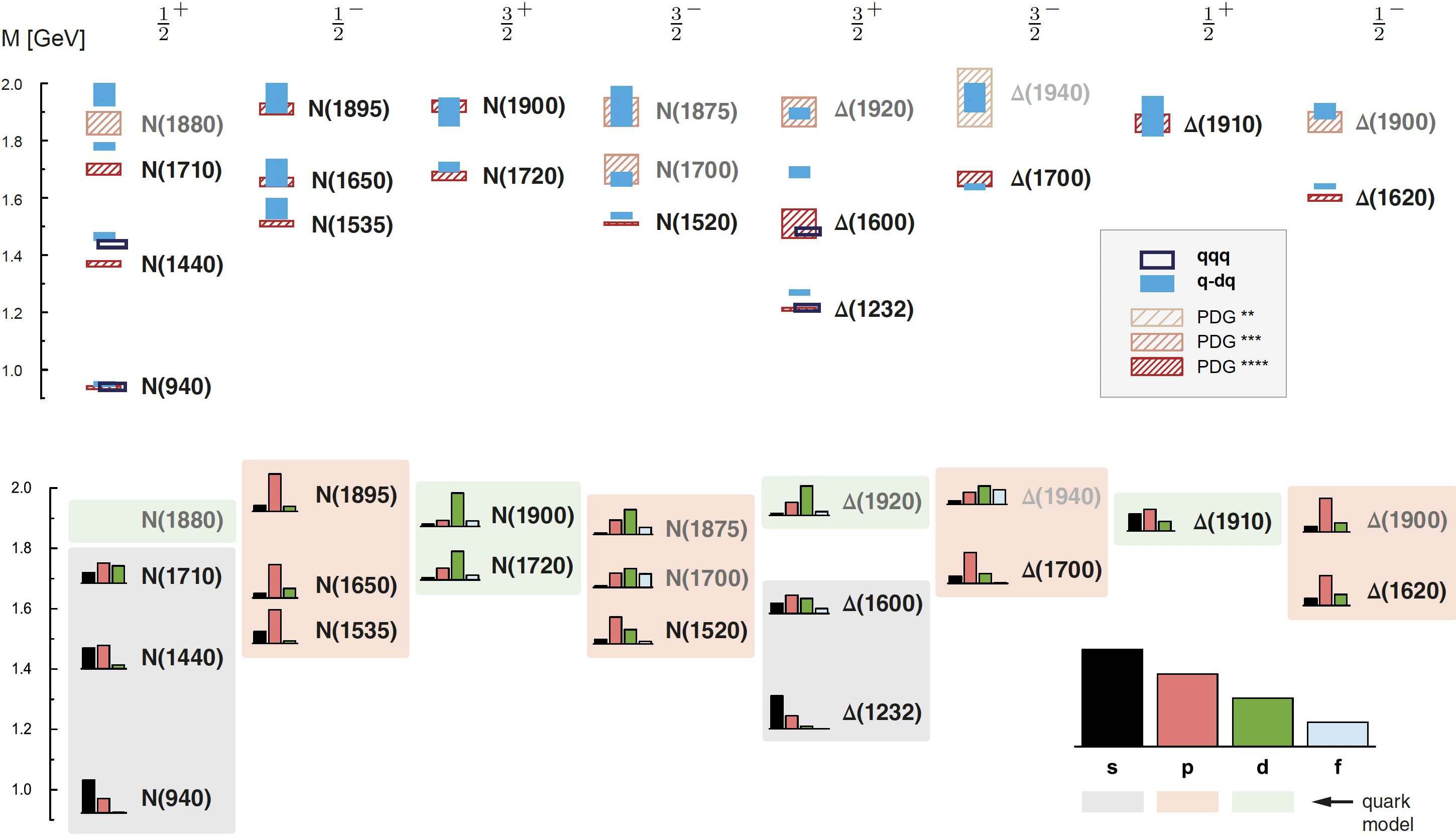}
                    \caption{Top: Light baryon spectrum from the three-quark and quark-diquark equations compared to the PDG~\cite{Eichmann:2016hgl}.
                             The three-quark and quark-diquark in the nucleon $1/2^+$ and $\Delta$-baryon $3/2^+$ channels show good agreement, 
                             and there is a 1:1 correspondence between the calculated and experimental spectrum. Bottom: Orbital angular momentum components
                             for the different states. The colored rectangles correspond to the orbital angular momentum assignments in the nonrelativistic 
                             quark model.}\label{fig:spectrum-qdq}
            \end{figure*}

  The top panel in Fig.~\ref{fig:spectrum-qdq} shows the resulting light-baryon excitation spectrum for $J=1/2^\pm$ and $J=3/2^\pm$ quantum numbers~\cite{Eichmann:2016hgl}.
  The open boxes are the three-quark results, the filled boxes the quark-diquark results, and 
  the hatched boxes are the PDG masses. One can see that the three-quark and quark-diquark results in the nucleon $1/2^+$ and $\Delta$-baryon $3/2^+$ channels
  are very similar. Apparently, diquark clustering is a good way to think about baryons! 
  
  For the remaining channels, we briefly need to discuss the diquark properties.
   The lightest ones are the scalar diquarks (with $J^P = 0^+$, isospin $I=0$ and masses around 800 MeV), followed by the axialvector diquarks 
   ($J^P = 1^+$, isospin $I=1$ and masses around 1~GeV), then
  pseudoscalar diquarks, vector diquarks, etc.
  It turns out that the scalar (axialvector) diquarks are closely linked with the pseudoscalar (vector) mesons; their BSEs differ only by a color factor 2.
  In the literature these are also often called `good' and `bad' diquarks, although  
  there is nothing good or bad about them: both scalar and axialvector diquarks contribute to the nucleon and Roper resonance,
  and for the $\Delta$ baryon only axialvector diquarks (the `bad' ones!) can contribute due to isospin.

  Quoting a famous spaghetti Western, 
  if scalar diquarks are `good' and axialvector diquarks `bad', then the higher-lying pseudoscalar and vector diquarks are indeed  `ugly'.
  These are the analogues of the more problematic scalar and axialvector mesons,
  whose rainbow-ladder masses come out too low. They also strongly affect the remaining 
  baryon channels in Fig.~\ref{fig:spectrum-qdq}, i.e., nucleon $1/2^-$ and $3/2^\pm$ and $\Delta$-baryon $3/2^-$ and $1/2^\pm$,
  whose masses also come out too low.
  In the three-quark approach, where  diquarks do not  appear explicitly,  there is nothing one can do about this except go beyond rainbow-ladder. 
  However, in the quark-diquark system one can tune the properties of the `ugly' diquarks: 
  One can simulate beyond rainbow-ladder effects by introducing one additional strength parameter, which is fixed in the meson sector
  and pushes up the axialvector mesons to their experimental ballpark.
  This also pushes up the baryon masses in these channels,
  which leads to the plot  in Fig.~\ref{fig:spectrum-qdq}. It shows  a 1:1 correspondence between the calculated
  and experimental spectrum, and there are no missing states. In total this gives two active parameters,
  the rainbow-ladder scale $\Lambda$ (fixed by $f_\pi$) and the diquark strength parameter (fixed by the $a_1/b_1$ mesons).
  This may be compared to quark models, where sometimes a dozen parameters are needed to describe the spectrum!
  
  Speaking of the quark model, in nonrelativistic quark models each baryon is usually identified  with 
  a definite orbital angular momentum $L$. The nucleon and $\Delta$-baryon have $L=0$, the $N(1535)$ and $N(1650)$ have $L=1$, and so on.
  Relativistically this is no longer so, because only the total angular momentum $J$ is a good quantum number
  while different $L$ components mix.
  For example, each tensor $\tau_i(p,q,P)$ in Eq.~\eqref{baryon-amp} 
  carries a specific value of $L$ in the baryon's rest frame, but the baryon is a superposition of all of them. 
  This is shown in the bottom of Fig.~\ref{fig:spectrum-qdq}: The nucleon and $\Delta(1232)$ carry $L=0$ ($s$-wave) 
  but also $L=1$ ($p$-wave) components, where the latter are relativistic effects. 
  In many cases  
  the leading components agree with the quark model identification,
  so the nonrelativistic quark model clearly shines through in the spectrum. 
  In general, however, relativity is quite important for the light baryon spectrum.

  A quark-diquark picture is also appropriate for hyperons containing strange quarks, 
  as well as for charmed baryons such as $\Sigma_c$ and $\Lambda_c$ which consist of two light quarks and one charm quark. 
  Denoting up and down by $n$, it is often assumed that charmed baryons form `planetary' systems made of a light $(nn)$ diquark that `orbits' around the heavier charm quark.
  However, the Faddeev calculations do not support this picture~\cite{Yin:2019bxe,Torcato:2023ijg}. Rather, the $(nc)n$ configurations are dominant and contribute about $60\%$ to the norm of the state,
  while the $(nn)c$ configurations contribute only $40\%$. Therefore, a charmed baryon actually spends most of its time in a $(nc)n$ configuration.

\section{Four-quark states}\label{sec:tetraquarks}

\subsection{Dynamical generation of resonances}\label{sec:resonances}

\begin{wrapfigure}[18]{r}{0.22\textwidth}
\vspace{-10mm}
\includegraphics[width=\linewidth]{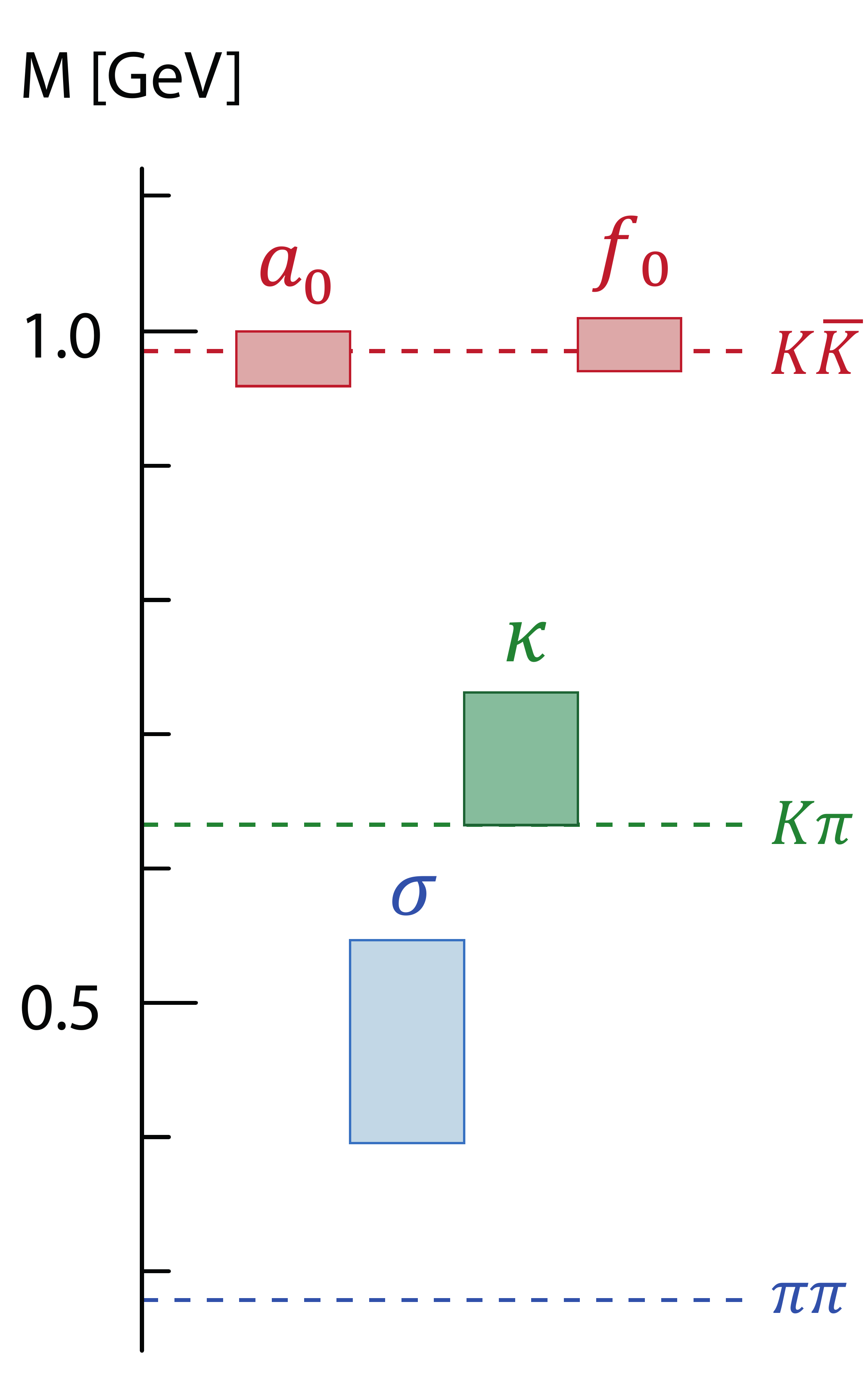} 
\caption{Light scalar meson spectrum from the PDG~\cite{Eichmann:2020oqt}.}
\label{fig:4q-light-sc}
\end{wrapfigure}

Given that diquarks appear to play  an important role for baryons, 
it is natural to ask whether diquark clustering also extends to exotic hadrons.
The oldest exotic meson candidate is the $\sigma/f_0(500)$ meson, 
 which is part of  the lightest scalar meson nonet with quantum numbers $J^{PC} = 0^{++}$ shown in Fig.~\ref{fig:4q-light-sc}.
 The $\sigma$ has several unusual properties.
 Viewed as  ordinary $q\bar{q}$ states, the $a_0$ and $\sigma$ should be approximately mass-degenerate,  
 like the $\rho$ and $\omega$ mesons in the $1^{--}$ vector channel, since both are made of light up and down quarks.
 The kaon-like $\kappa/K_0^\ast(700)$ with one strange quark should  be heavier, followed by the $f_0(980)$ which is the $s\bar{s}$ state. 
 However, this is not what we see in Fig.~\ref{fig:4q-light-sc}: Instead, the $a_0$ and $f_0$ are mass-degenerate and the heaviest states,
 which does not make any sense.
 In the nonrelativistic quark model, the light scalar nonet also carries orbital angular momentum $L=1$ 
 and should thus be considerably heavier than the pseudoscalar and vector mesons.
 Furthermore, the $\sigma$  has a huge decay width compared to other $q\bar{q}$ states, which makes it hard to see in experiments
 and has led to a long controversy regarding its status in the PDG~\cite{Pelaez:2015qba}.

 A possible explanation is that these states could be four-quark states in the form of diquark-antidiquark clusters $(qq)(\bar{q}\bar{q})$~\cite{Jaffe:1976ig,tHooft:2008rus}.
 According to Eq.~\eqref{color-baryon}, a scalar $\overline{\mathbf{3}}$ diquark can combine with a scalar $\mathbf{3}$ antidiquark
 to form a color singlet. This produces again a nonet of states, but now the mass ordering is reversed: The $\sigma$ is the lightest state
 because it is made of four light quarks, while the $\kappa$ contains one strange quark and the $a_0$ and $f_0$ two, which explains the mass ordering.
 It would also explain the large decay widths, since the $\sigma$ can simply fall apart to produce two pions without exchanging gluons.
 Alternatively, the observation that the $a_0$ and $f_0$ sit very close to the $K\bar{K}$ threshold points toward a molecular picture $(q\bar{q})(q\bar{q})$~\cite{Weinstein:1990gu}.
 In any case, there seems to be a general consensus by now that the light scalar mesons are not $q\bar{q}$ but rather
 four-quark states~\cite{Pelaez:2015qba}.

              \begin{figure*}[t]
                    \centering
                    \includegraphics[width=1.0\textwidth]{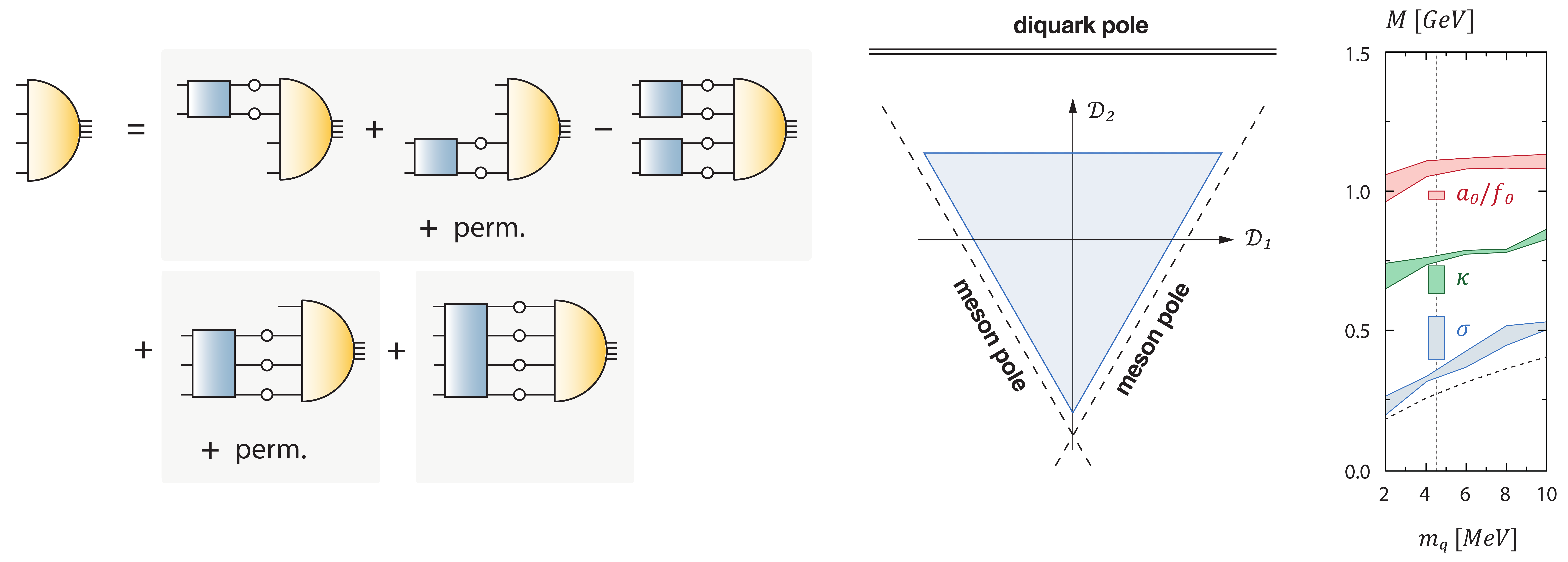}
                    \caption{Left: Four-quark Bethe-Salpeter equation with two-, three- and four-body kernels.
                             Center: Mandelstam plane where the meson-meson and diquark-antidiquark poles become visible.
                             The colored triangle is the integration region in the BSE. Right: Light scalar meson spectrum
                             from the four-quark equation in the vicinity of the light current-quark mass~\cite{Eichmann:2015cra};
                             the boxes are the PDG masses.
                     }\label{fig:4q-bse}
            \end{figure*} 
 
 To study the light scalar mesons with functional methods, one should solve the four-quark Bethe-Salpeter  (or Faddeev-Yakubowski) equation
 shown in the left of Fig.~\ref{fig:4q-bse}. In this case, one has permutations of two-body irreducible kernels (minus subtraction terms to avoid overcounting)
 plus three- and four-body kernels. Keeping only the two-body kernels, the situation is worse compared to baryons  
 because the wave function (the four-body analogue of Eqs.~\eqref{pion-amp} and~\eqref{baryon-amp})  depends on four independent momenta $p_1, \dots p_4$. This
  results in hundreds of Dirac tensors, whose dressing functions $f_i$ depend on nine kinematic variables (plus the total onshell momentum variable $P^2$).
 The resulting BSE is an eigenvalue equation whose kernel matrix has a dimension of about $10^{13}$ --- this is impossible to solve even with the best computers
 that are  available today.
 
 A very useful way to construct approximations is to group these variables in multiplets of the permutation group $S_4$, 
 because this allows one to switch off groups of variables without breaking the symmetries of the system.
 It turns out that one can arrange the $1+2+3+3=9$ variables into a singlet $\mS_0$, a doublet $\mD$ and two triplets.
 The most important variables are the singlet and doublet. The singlet is the symmetric variable $\mS_0 \sim p_1^2 + p_2^2 + p_3^2 + p_4^2$
 and carries the scale. The doublet forms the Mandelstam plane  in the center panel of Fig.~\ref{fig:4q-bse}, which shows the internal two-body poles.
 If we label the quarks in the $qq\bar{q}\bar{q}$ wave function by 1, 2 and the antiquarks by 3, 4, then the system can form internal diquark-antidiquark clusters, (12)(34),
 and two meson-meson clusters, (13)(24) and (14)(23). These are dynamically generated in the four-quark BSE and show up as poles in the Mandelstam plane.
 Because the pions are much lighter than the diquarks, the pion poles are much closer to the integration region (the colored triangle)
 than the diquark poles. Thus, the four-body system is dominated by the pions: If the four-body BSE is solved keeping the variable $\mS_0$ only,
 the resulting $\sigma$ mass is about 1500 MeV; the two triplets only have subleading effects; but if the doublet $\mD$ is included,
 the mass jumps down to $\sim 350$ MeV~\cite{Eichmann:2015cra}. The resulting mass pattern for the light scalar mesons qualitatively agrees with the experimental spectrum,
 as  shown in the right panel of Fig.~\ref{fig:4q-bse}
 in the vicinity of the light current-quark mass.
 
 What happened here is quite interesting: One starts out with four quarks which interact by gluons. However, in the solution process
 the four-quark wave function dynamically generates pion poles, which produce a resonance mechanism. Thus, the resulting $\sigma$ meson
 is at the same time a four-quark state \textit{and} a $\pi\pi$ resonance! This also explains why it is so light, since the dominant internal clusters
 are not the diquarks but rather the pions, the pseudoscalar Goldstone bosons in QCD.
 So, even though the underlying mechanism --- the dynamical generation of two-body clusters --- is similar as for baryons,
 the physical consequences are very different.
 
 This mechanism also builds a bridge to  effective field theories, where two-body BSEs and scattering equations
 are solved using hadronic degrees of freedom~\cite{Guo:2017jvc}.
 To make an analogy with civil engineering, suppose the hadronic degrees of freedom  from the effective Lagrangian 
 live on the first floor. Solving hadronic BSEs, these dynamically create resonances on the second floor.
 In the four-quark equation, we take the elevator and go a level deeper:
 The fundamental degrees of freedom are the quarks and gluons in the basement.
 Even though they are not directly observable, they keep the building running:
 They dynamically create hadrons on the first floor, 
 which may  dynamically create further resonances on the second floor. 
 In the four-quark equation all this happens automatically without making assumptions 
 (apart from practical approximations to make the numerical solutions feasible).

 The remaining question concerns the $q\bar{q}$ contributions to the $\sigma$.
 Even though the $\sigma$ resides on the second floor, it also has an office on the first floor since it may be a superposition of $qq\bar{q}\bar{q}$ and $q\bar{q}$ components.
 This can be studied by coupling the four-quark BSE with the $q\bar{q}$ BSE, from where one finds
 that the $q\bar{q}$ contributions are indeed rather small~\cite{Santowsky:2020pwd}.

              \begin{figure*}[t]
                    \centering
                    \includegraphics[width=0.9\textwidth]{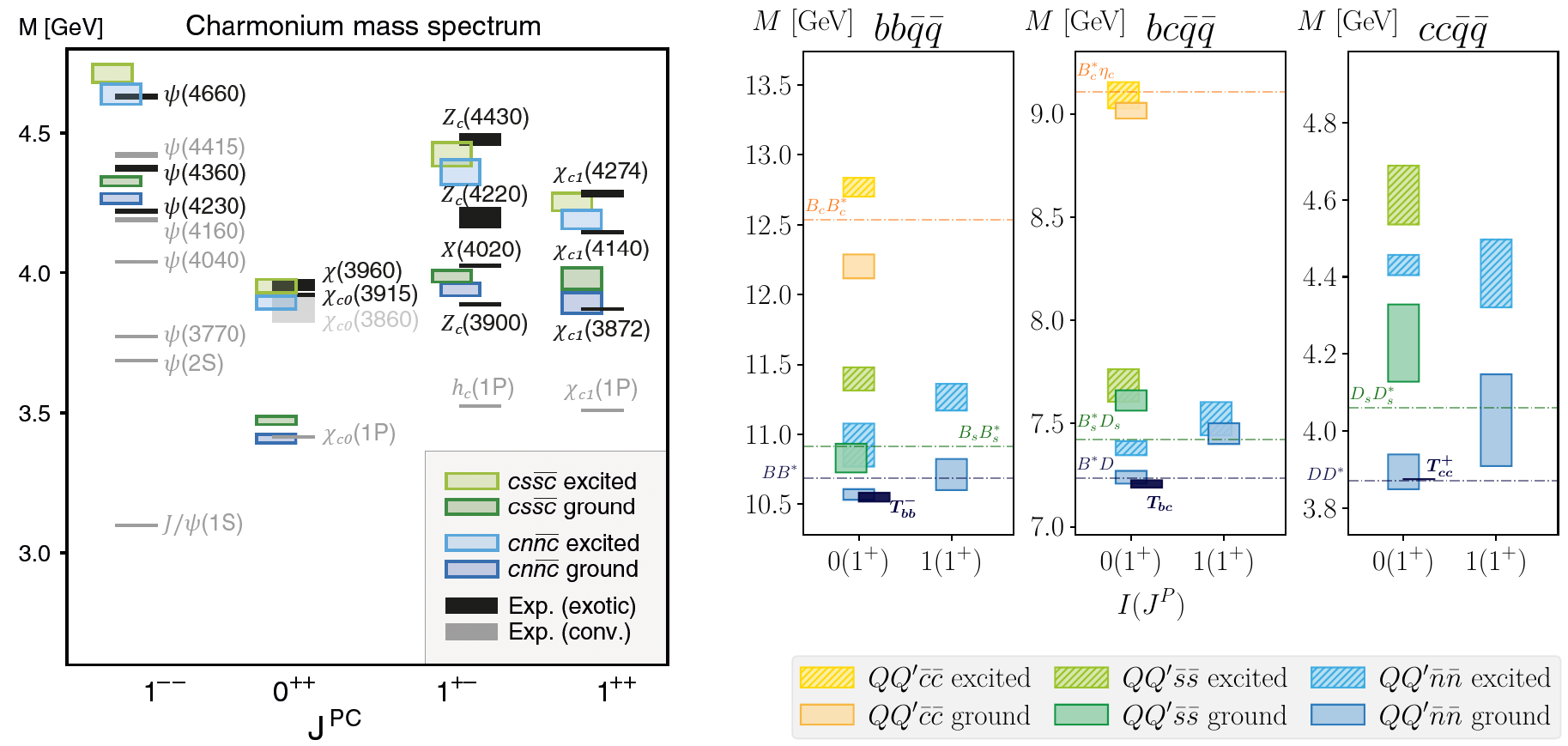}
                    \caption{Left: Hidden-charm spectrum. The gray boxes are the conventional $c\bar{c}$ states from the PDG, and the black boxes
                             are the exotic states. The colored boxes are the results from the four-quark equation~\cite{Hoffer:2024alv}.
                             Right: Open-flavor spectrum for $J^P = 1^+$ and quark content $bb\bar{q}\bar{q}$, $bc\bar{q}\bar{q}$ and $cc\bar{q}\bar{q}$.
                             The levels in black are the experimental $T_{cc}^+$ mass and the theory predictions for its partners.
                             The colored boxes are the four-quark results~\cite{Hoffer:2024fgm}.}\label{fig:4q-heavy}
            \end{figure*}

\subsection{Heavy exotics}

Compared to light mesons, where relativity and chiral symmetry are important,
the heavy-quark sector is a much cleaner environment for studying exotics.
Here  the success of the non-relativistic quark model provides a
basic measure for distinguishing `ordinary' from `exotic' mesons, namely through
the deviation from the quark-model expectations.
An abundance of new data in the charm and bottom region has established
the existence of four-quark states~\cite{Chen:2016qju,Lebed:2016hpi,Esposito:2016noz,Ali:2017jda,Olsen:2017bmm,Guo:2017jvc,Liu:2019zoy,Brambilla:2019esw}.
Many of them are close to meson-meson thresholds,
which points towards at a molecular nature. 
In the following we consider two types of exotic mesons:
\begin{itemize}
\item  \textit{Hidden-flavor states} of the type  $Q\bar{Q}q\bar{q}$, where $Q = c, b$ is a heavy quark
       and $q = u, d, s$. If these states carry isospin $I=0$, they can still mix with ordinary $Q\bar{Q}$ mesons.
       The most prominent example is the $\chi_{c1}(3872)$ with quantum numbers $I(J^{PC}) = 0(1^{++})$,
       which was first reported by Belle in 2003~\cite{Belle:2003nnu}. 
       It has several properties that do not fit into a conventional charmonium picture, like
       its mass which is indistinguishable from the $D^0 \bar{D}^{\ast 0}$ threshold,
        its narrow width ($<1.2$ MeV) which is much smaller than potential-model predictions for the first excited
        $c\bar{c}$ state, and its unusual decay modes.
        For other exotic candidates like the $\psi(4230)$ in the $1^{--}$ vector channel,
        their exotic assignment is mainly due to the overpopulation of the $1^{--}$ channel with regard to well-established  $c\bar{c}$ states.
        The $Z_c(3900)$ and $Z_c(4430)$ with $J^{PC} = 1^{+-}$, on the other hand, carry charge and are thus manifestly exotic
        since their minimal quark content is $c\bar{c}u\bar{d}$.

\item \textit{Open-flavor states} of the type $QQ\bar{q}\bar{q}$, which are manifestly exotic and cannot be made of $q\bar{q}$.
      So far, only one state has been found experimentally, the $T_{cc}(3875)^+$ with quark content $cc\bar{u}\bar{d}$,
      which is extremely close to the $D^0 D^{\ast +}$ threshold~\cite{LHCb:2021vvq}.
      If the masses of the heavy quarks are sufficiently large and those of the light quarks sufficiently small,
      such states are expected to form deeply bound states~\cite{Francis:2016hui,Eichten:2017ffp}, like the 
      $T_{bb}^-$ made of $bb\bar{u}\bar{d}$ which is predicted by many theoretical calculations.

\end{itemize}
Meanwhile also states with one heavy quark ($Q\bar{q}q\bar{q}$) and four heavy quarks ($Q\bar{Q}Q\bar{Q}$) have been found with LHCb,
so the list of heavy exotics keeps ever growing~\cite{LHCb-FIGURE-2021-001-report}.

What can the four-quark equation say about heavy exotics? In principle, the approach discussed in Sec.~\ref{sec:resonances} can be taken over without modifications.
For hidden-charm states, the three sides of the triangle in Fig.~\ref{fig:4q-bse} correspond to the three possible two-body clusters in the system:
diquark-antidiquark $(Qq)(\bar{Q}\bar{q})$, meson molecule $(Q\bar{q})(q\bar{Q})$, or hadroquarkonium $(Q\bar{Q})(q\bar{q})$.
Take for example the $\chi_{c1}(3872)$ made of $c\bar{c}n\bar{n}$, where $n=u,d$ are the light up and down quarks.
Here the lightest clusters are a meson-molecule configuration $(c\bar{n})(n\bar{c})$ which is $D\bar{D}^\ast$,  a hadroquarkonium component $(c\bar{c})(n\bar{n})$ which is $J/\psi \omega$, and a diquark-antidiquark $(cn)(\bar{c}\bar{n})$ configuration which is substantially heavier.

The resulting spectrum is shown in the left panel of Fig.~\ref{fig:4q-heavy}. The gray levels are the conventional $c\bar{c}$ states from experiment,
the black levels the exotic candidates, and the colored boxes are the BSE results for $cn\bar{n}\bar{c}$ and $cs\bar{s}\bar{c}$ states 
(ground state and first excited state each). 
In general, the proximity of the $cn\bar{n}\bar{c}$ and $cs\bar{s}\bar{c}$ states
leads to a densely populated spectrum. 
One can see that the $\chi_{c1}(3872)$ in the $1^{++}$ channel is nicely reproduced (note, however, that the backcoupling of the $c\bar{c}$
components is not yet included here).  
For the $1^{--}$ and $1^{+-}$ states an identification is also possible. An exception is the $0^{++}$ channel, where the four-body results are close to the 
ordinary $c\bar{c}$ state. Concerning the internal structure of these states, the $1^{--}$, $0^{++}$ and $1^{++}$ channels are strongly dominated
by meson-molecule configurations (like $D\bar{D}^\ast$ for the $\chi_{c1}(3872)$), while in the $1^{+-}$ channel they are mixture 
of meson-molecule and hadrocharmonium components, where the latter is dominant. The diquark components are almost negligible in all cases.

For open-charm states the situation is quite different. Here the two-body clusters are diquark-antidiquark $(QQ)(\bar{q}\bar{q})$
and two degenerate meson-meson configurations $(Q\bar{q})(Q\bar{q})$. The resulting spectra for 
$bb\bar{q}\bar{q}$, $bc\bar{q}\bar{q}$ and $cc\bar{q}\bar{q}$ states with $J^P = 1^+$  are shown in the right panels of Fig.~\ref{fig:4q-heavy}.
In the rightmost plot, the black level is the experimental mass of the $T_{cc}^+$, while its partners in the other two plots are theory predictions.
Also here, the BSE results nicely match the ground states. The internal structure, however, turns out to be very different.
The $T_{cc}^+$ is almost entirely dominated by the molecular $D D^{\ast}$ component due to its proximity to the threshold. 
Its bottom partner $T_{bb}^-$, on the other hand, is a mixture of everything with a dominant diquark-antidiquark component.

One can see that the strength of the four-body equation is its ability to \textit{predict} the dominant two-body clusters. 
This is different from model calculations, where one settles on one specific component (diquark-antidiquark, meson molecule, hadroquarkonium) from the start
and makes predictions for that component. 
However, the size of these components may be very different on a case-by-case basis or even a complicated mixture,
and it is a dynamical question which component is the leading one.
In the four-body BSE these components are  dynamically generated  from the quarks and gluons in the equation without prior assumptions.

              \begin{figure*}[t]
                    \centering
                    \includegraphics[width=1\textwidth]{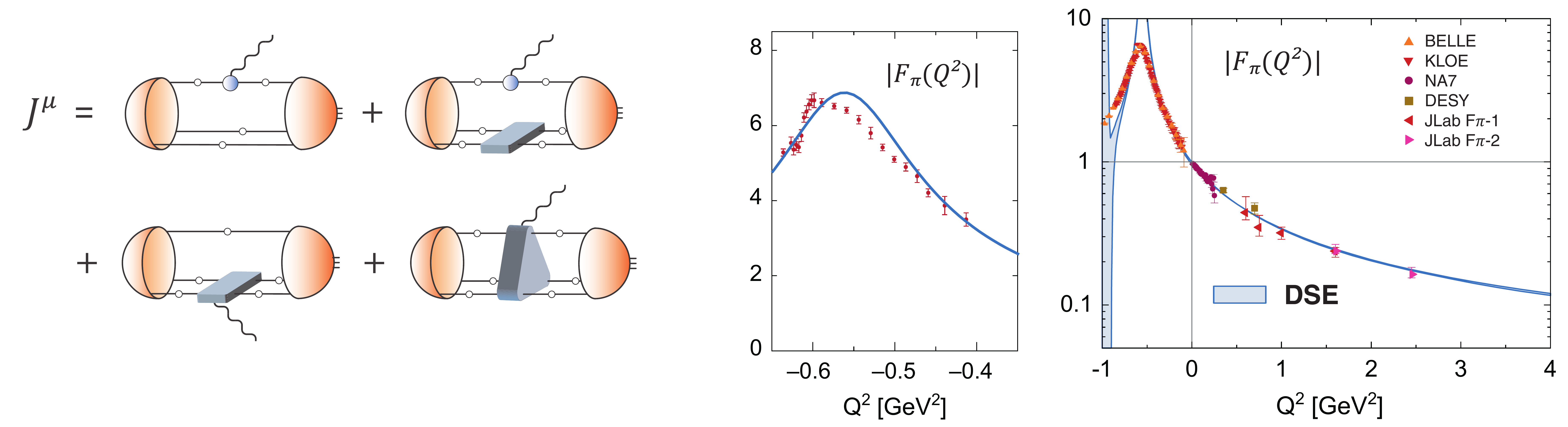}
                    \caption{Left: Current matrix element for a three-quark system. The photon couples to the quarks and to the
                             two- and three-quark irreducible kernels. Center and right: Electromagnetic pion form factor
                             in the spacelike and timelike region, with DSE/BSE results  compared to experimental data.
                             The right panel shows the rainbow-ladder result, where the timelike $\rho$-meson pole is visible~\cite{Eichmann:2019tjk}.
                             The center panel is the result of  a calculation beyond rainbow-ladder including the $\pi\pi$ channel;
                             here the pole is shifted onto the second Riemann sheet and only the bump is visible on the real axis~\cite{Miramontes:2021xgn}.
                             }\label{fig:ffs}
            \end{figure*} 

\section{Hadron structure}\label{a}

So far we have talked about hadron \textit{spectroscopy} with functional methods. To recapitulate, the strategy is to solve two-, three- and four-body
BSEs (Figs.~\ref{fig:quark-dse}, \ref{fig:faddeev}, \ref{fig:qdq} and~\ref{fig:4q-bse}), whose ingredients are QCD's $n$-point functions 
which must be calculated beforehand. The outcomes of the BSEs are the masses, i.e., the excitation spectra for the various $J^{P(C)}$ channels,
and the  hadron wave functions, i.e., their Bethe-Salpeter amplitudes. With the wave functions at hand, one can  go further and
calculate hadron \textit{structure} properties such as form factors, parton distributions and scattering amplitudes, which are encoded in the hadron matrix elements.

These matrix elements can be derived in analogy to the derivation of the BSE itself  in Sec.~\ref{sec:bse}.
For example, to obtain an electromagnetic nucleon-to-resonance transition matrix element ($\gamma^\ast N \to N^\ast$), 
one starts from the quark six-point function
(the three-quark analogue of the $q\bar{q}$ four-point function in Fig.~\ref{fig:sc-eq}), couples a photon to it, 
and inserts a complete set of states to the left and right of the current insertion.
This produces all possible baryon poles, once on the left and once on the right of the photon insertion.
The matrix element is then the residue at the ($N$, $N^\ast$) double pole~\cite{Eichmann:2011ec}.
Once again, it does not matter if the poles are real or complex, i.e, whether one deals with bound states or resonances,
because the current matrix element is simply the residue at the pole.
The result is shown in the left panel of Fig.~\ref{fig:ffs}: The photon couples to the quarks
and to the two- and three-body irreducible kernels.
Note that this is an exact expression in QFT. In a rainbow-ladder truncation only the first two diagrams survive,
because the photon cannot couple to a gluon and the three-body kernel is omitted.

The only new ingredient  is then the quark-photon vertex, which is the $q\bar{q}$ four-point function $\mathbf{G}$ in Fig.~\ref{fig:sc-eq} 
with one $q\bar{q}$ pair  contracted to form a photon. The scattering equation for $\mathbf{G}$ then immediately yields
an inhomogeneous BSE for the vertex, which can be solved with the same input as outlined earlier and without further approximations~\cite{Maris:1999bh}.
The BSE for the quark-photon vertex has some interesting features. Because $\mathbf{G}$ contains all possible meson poles,
the quark-photon vertex must pick up all meson poles with the same quantum numbers of the photon.
These are the vector mesons with $J^P = 1^{--}$.
Thus, when solving the vertex BSE, these poles are dynamically generated in the solution.
This is the underlying origin of vector-meson dominance, which dictates the shape of a spacelike electromagnetic form factor at low $Q^2$:
On the timelike side, the electromagnetic form factors of hadrons must have vector-meson poles.

As an example, the right panel in Fig.~\ref{fig:ffs} shows the pion electromagnetic form factor, which has a pole at $Q^2 = -m_\rho^2$.
This is an automatic feature when calculating the form factor and its ingredients systematically from their DSEs and BSEs.
The normalization $F_\pi(0) = 1$ is also automatic as long as the BSE kernel satisfies the vector Ward-Takahashi identity.
In reality these poles are bumps, because the $\rho$-meson is a resonance with a pole on the second Riemann sheet.
The rainbow-ladder truncation generates a pole on the real axis corresponding to a bound state, but
if one includes intermediate $\pi\pi$ channels in the BSE kernel, the $\rho$ pole moves into the complex plane
on the second Riemann sheet and the $\rho$ meson becomes a resonance,
as shown in the center panel.

Based on this, many form factor calculations have been done over the last decades, see e.g.~\cite{Eichmann:2016yit,Barabanov:2020jvn,Segovia:2015hra} and references therein: 
meson electromagnetic and transition form factors, baryon elastic and  transition form factors,
axial and gravitational form factors, meson two-photon transition form factors that enter in the muon $g-2$ problem, and so on. 
For baryons one finds similar conclusions as discussed in Sec.~\ref{sec:baryons}:
The three-quark and quark-diquark results are very similar, and discrepancies between the two approaches are rather small. 
Discrepancies with experimental results mainly appear at small $Q^2$, where meson-cloud effects become important.
These are not yet captured by a rainbow-ladder truncation but, as we just mentioned, work in this direction is underway.
Also a range of hadron structure calculations are available, e.g., for pion and kaon parton distribution functions~\cite{Aguilar:2019teb},
and scattering amplitudes such as $\pi\pi \to \pi\pi$, $\pi\gamma^\ast\to\pi\pi$ and $N\gamma^\ast \to N\gamma^\ast$ have been investigated as well~\cite{Cotanch:2002vj,Eichmann:2012mp,Eichmann:2016yit}.  

\newpage

\section{Conclusions and outlook}\label{sec:conclusions}

  We have seen that functional methods provide a systematic way to calculate hadron properties
  from quarks and gluons, the fundamental degrees of freedom of QCD.
  Despite 25+ years of work, the approach is still at a comparatively early stage
  given that the most hadron physics applications so far have employed
  the rainbow-ladder truncation, or minimal extensions of it, or simpler models.
  To systematically go beyond the status quo towards precision calculations, one needs to calculate QCD's 
  $n$-point functions beyond the quark propagator which go into the BSE kernel:
  the gluon propagator, quark-gluon vertex, three- and four-gluon vertex, and so on.
  There are ongoing efforts in these directions using the
   Dyson-Schwinger equations~\cite{Huber:2020keu,Horak:2021pfr,Eichmann:2021zuv,Ferreira:2023fva} shown in Fig.~\ref{fig:dses},
  the functional renormalization group~\cite{Cyrol:2017ewj,Dupuis:2020fhh}, and gauge-fixed lattice QCD~\cite{Falcao:2020vyr,Kizilersu:2021jen,Aguilar:2024dlv,Pinto-Gomez:2024mrk}.
  A nice example is the glueball spectrum in Yang-Mills theory, i.e., QCD without quarks:
  Once the underlying $n$-point functions are calculated from their DSEs,
  the resulting glueball spectrum agrees with lattice QCD~\cite{Huber:2021yfy}.
  In the next years we will probably see similar advances  along these lines
  towards QCD with dynamical quarks.
  
  A parallel avenue is to expand the pool of observables that can be calculated with functional methods.
  As we have seen, applications for meson  and baryon spectroscopy (including four-quark states) are reasonably advanced by now.
  The same can be said  about form factors and parton distributions, as well as 
  for applications towards finite temperature and density~\cite{Fischer:2018sdj}.
  The next frontier are exotic quantum numbers, many-quark states, 
  scattering amplitudes and weak interactions of hadrons,  generalized parton distributions and transverse momentum distributions, or fragmentation functions and hadronization.
  Exploratory studies in several of these directions are already underway,
  and as the functional methods community is  growing bigger  we may see many more results in the future.

 \begin{ack}[Acknowledgments]%
This work was supported by the Austrian Science Fund FWF under grant number 10.55776/PAT2089624 
and by the Portuguese Science fund FCT 
under grant number CERN/FIS-PAR/0023/2021.
This work contributes to the aims of the USDOE ExoHad Topical Collaboration, contract DE-SC0023598.

\end{ack}

\bibliographystyle{Numbered-Style} 
\bibliography{reference}

\end{document}